\documentclass[runningheads]{llncs}
\usepackage[T1]{fontenc}
\usepackage{graphicx}
\usepackage{svg}
\usepackage{amsmath}
\usepackage{amssymb}
\usepackage{graphicx}
\usepackage{semantic}
\usepackage{xspace}
\usepackage{placeins}
\usepackage{subcaption}
\usepackage{mathtools}
\usepackage{comment}
\usepackage{multirow}
\usepackage{hyperref}
\usepackage{cleveref}
\usepackage{xcolor}
\usepackage{orcidlink}
\usepackage{bbding}
\usepackage{standalone}
\usepackage[disable]{todonotes}
\usepackage{xcolor}
\usepackage{orcidlink}
\usepackage{bbding}
\usepackage{bm}
\usepackage{wrapfig}
\usepackage{currfile}
\newcommand{\Loc}{\ensuremath{\mathit{Loc}}\xspace}
\newcommand{\Var}{\ensuremath{\mathit{Var}}\xspace}

\newcommand{\VarInner}{\ensuremath{\mathit{Var}_\mathit{output}}\xspace}
\newcommand{\VarInnerI}[1]{\ensuremath{\mathit{Var}_\mathit{output, #1}}\xspace}
\newcommand{\VarOuter}{\ensuremath{\mathit{Var}_\mathit{input}}\xspace}

\newcommand{\Inv}{\ensuremath{\mathit{Inv}}\xspace}
\newcommand{\Init}{\ensuremath{\mathit{Init}}\xspace}
\newcommand{\Flow}{\ensuremath{\mathit{Flow}}\xspace}

\newcommand{\guard}{\ensuremath{\mathit{g}}\xspace}
\newcommand{\reset}{\ensuremath{\mathit{r}}\xspace}

\newcommand{\state}{\ensuremath{\sigma}\xspace}

\newcommand{\Sync}{\ensuremath{\mathit{Sync}}\xspace}
\newcommand{\SyncS}{\ensuremath{\mathit{Sync}_{\mathit{send}}}\xspace}
\newcommand{\SyncR}{\ensuremath{\mathit{Sync}_{\mathit{receive}}}\xspace}
\newcommand{\LAC}{\ensuremath{\mathit{LHAC}}\xspace}
\newcommand{\lacsyindex}{sync\xspace}
\newcommand{\LACsync}{\ensuremath{\mathit{LHAC}}\lacsyindex}

\newcommand{\LACsyncInstanceI}[1]{\ensuremath{\mathcal{L}_{\lacsyindex, #1}}\xspace}

\newcommand{\out}{\ensuremath{\mathit{out}}\xspace}
\newcommand{\innone}{\ensuremath{\mathit{in}}\xspace}
\newcommand{\inone}{\ensuremath{\mathit{in}_1}\xspace}
\newcommand{\intwo}{\ensuremath{\mathit{in}_2}\xspace}

\newcommand{\fail}{\ensuremath{\mathit{fail}}\xspace}
\newcommand{\repair}{\ensuremath{\mathit{repair}}\xspace}

\newcommand{\intermediateaut}{intermediate automaton\xspace}

\newcommand{\realyst}{\textsc{RealySt}\xspace}
\newcommand{\modest}{\textsc{ModestToolset}\xspace}
\newcommand{\uppaal}{\textsc{Uppaal}\xspace}

\newcommand{\LHACSetSync}{\ensuremath{\mathbf{L}_{\mathit{comp}}}\xspace}

\newcommand{\MappingSet}{\ensuremath{\mathcal{M}}\xspace}

\newcommand{\MappingI}[1]{\ensuremath{m_{#1}}\xspace}
\newcommand{\MappingComp}{\ensuremath{m_{\mathit{comp}}}\xspace}
\newcommand{\variable}{\ensuremath{\mathit{name}}\xspace}
\newcommand{\receiveSet}{\ensuremath{\mathit{Rcv}}\xspace}
\newcommand{\sender}{\ensuremath{\mathit{snd}}\xspace}

\newcommand{\F}{\mathcal{F}}

\newcommand{\N}{\mathbb{N}}
\newcommand{\R}{\mathbb{R}}

\newcommand{\Rpz}{\mathbb{R}_{\geq 0}}

\newcommand{\States}{\Sigma}

\newcommand{\suchthat}{\,|\,}
\newcommand{\support}{\textit{supp}}
\newcommand{\Distr}{\textit{Dist}}
\renewcommand{\H}{\mathcal{H}\xspace}

\newcommand{\Lab}{\mathit{Lab}}
\newcommand{\Edge}{\mathit{Edge}}

\newcommand{\kernel}{\ensuremath{\Psi}}

\newcommand{\realisierungen}{\mathcal{R}}

\newcommand{\win}{\textit{w}}

\newcommand{\HEE}{\H^{{*}{*}}}

\newcommand{\HEEdecomp}{\HEE_{\textit{dec}}}
\newcommand{\auxwatch}{c}

\newcommand{\dep}{\mathit{len}}
\newcommand{\resetKernel}{\ensuremath{\kernel^R}}

\newcommand{\FlowE}{\Flow^{*}}

\newcommand{\FlowEdecomp}{\FlowE_{\textit{dec}}}

\newcommand{\EdgeE}{\Edge^{*}}

\newcommand{\EdgeEdecomp}{\EdgeE_{\textit{dec}}}
\newcommand{\VarE}{\Var^{{*}}}

\newcommand{\VarEdecomp}{\VarE_{\textit{dec}}}
\newcommand{\InvE}{\Inv^{*}}

\newcommand{\InvEdecomp}{\InvE_{\textit{dec}}}
\newcommand{\InitE}{\Init^{*}}

\newcommand{\InitEdecomp}{\InitE_{\textit{dec}}}

\newcommand{\gE}{g^{*}}

\newcommand{\gEdecomp}{\gE_{\textit{dec}}}

\newcommand{\rE}{r^{*}}

\newcommand{\rEdecomp}{\rE_{\textit{dec}}}

\newcommand{\Jumps}{E_{\States}}

\newcommand{\stopwatchDecomposed}[1]{\auxwatch_{#1}}

\newcommand{\ass}{\nu}

\newcommand{\hawk}{HAwK\xspace}
\newcommand{\hawksync}{\ensuremath{\mathit{HAwK}}\lacsyindex}
\newcommand{\hawksyncinstance}{\ensuremath{\mathcal{A}_{\mathit{sync}}}\xspace}
\newcommand{\hawksyncinstanceI}[1]{\ensuremath{\mathcal{A}_{\mathit{sync, #1}}}\xspace}
\newcommand{\HawkIAInstance}{\ensuremath{\hawksyncinstanceI{\mathcal{I}}}\xspace}
\newcommand{\HAWKSetNew}{\ensuremath{\mathbf{A}_{\mathit{new}}}\xspace}
\newcommand{\HawkIAInstanceNew}{\hawksyncinstanceI{\mathit{new}}}
\newcommand{\HAWKSetSync}{\ensuremath{\mathbf{A}_{\mathit{comp}}}\xspace}
\newcommand{\HAWKSet}{\ensuremath{\mathbf{A}}\xspace}
\newcommand{\HAWKSetComposed}{\ensuremath{\mathbf{A}_{\mathcal{I}}}\xspace}

\usepackage{xcolor-solarized}
\definecolor{solarizedblue}{RGB}{38 139 210}
\definecolor{solarizedpink}{RGB}{211, 54, 130}
\definecolor{highlight}{RGB}{43, 85, 128}
\definecolor{solarizedyellow}{RGB}{181 137   0}
\definecolor{coolyellow}{RGB}{191, 128, 15}
\definecolor{wwublue}{named}{highlight}
\definecolor{wwugray}{RGB}{62, 62, 59}
\definecolor{lightblue}{rgb}{0.6,0.9,1}
\definecolor{lightred}{rgb}{1,0.8,0.8}
\definecolor{rosa}{rgb}{0.9,0.6,0.6}
\definecolor{green}{rgb}{0.2,0.6,0.2}

\definecolor{steps}{named}{solarized-magenta}
\definecolor{simulink2sha}{named}{coolyellow}
\definecolor{uncertain}{named}{solarized-blue}
\definecolor{formal}{named}{solarized-green}
\definecolor{quantitative}{named}{solarized-violet}

\definecolor{send}{rgb}{0, 0.749, 0.078}
\definecolor{receive}{rgb}{0.008, 0.518, 0.812}
\definecolor{sync}{rgb}{0,0,0}
\definecolor{continuous}{rgb}{0.858, 0.188, 0.478}
\definecolor{discrete}{rgb}{0.858, 0.188, 0.478}
\definecolor{stochastic}{rgb}{0.769, 0.02, 0.624}
\definecolor{urgent}{rgb}{0,0,0} 
\definecolor{outer}{rgb}{0,0, 0}
\definecolor{inner}{rgb}{0,0,0}
\usepackage{tikz}
\usetikzlibrary{backgrounds}
\usetikzlibrary{positioning}
\usetikzlibrary{calc}
\usetikzlibrary{shapes,arrows}
\usetikzlibrary{patterns,hobby}
\usetikzlibrary{shadows,fit}
\usetikzlibrary{matrix}
\usetikzlibrary{decorations}

\newcommand{\innervar}[1]{\textcolor{inner}{#1}}
\newcommand{\outervar}[1]{\textcolor{outer}{#1}}
\newcommand{\send}[1]{\textcolor{send}{#1}}
\newcommand{\receive}[1]{\textcolor{receive}{#1}}

\newcommand{\randomclock}[1]{\textcolor{stochastic}{#1}}
\newcommand{\randomclockH}[1]{\textcolor{solarized-cyan}{#1}}

\newcommand{\simulink}[1]{\texttt{#1}}

\tikzset{subsystem/.style={rectangle, align = center, draw, fill=gray!10, minimum width = 0.8cm, minimum height = 0.4cm, font=\footnotesize, thick} }
\tikzset{blocklabel/.style={font=\tiny, align = center, inner sep = 1pt}}
\tikzset{dot/.style={circle,draw,fill, minimum width = 0.5ex, minimum height = 0.5ex,inner sep=0pt}}
\tikzset{port/.style={rectangle, rounded corners, align = center, draw, fill=none, minimum width = 0.6cm, minimum height = 0.3cm, font=\scriptsize, thick} }
\tikzset{goto/.style={signal, align = center, draw, fill=none, minimum width = 0.8cm, minimum height = 0.4cm, font = \bfseries\scriptsize, thick} }
\tikzset{location/.style={rounded corners, rectangle, draw, fill = none, align=center,font = \footnotesize, thick}}
\tikzset{label/.style={fill = none, align=center,font = \footnotesize}}
\tikzset{discrete/.style={color = discrete}}
\tikzset{continuous/.style={color = continuous}}
\tikzset{stochastic/.style={color = stochastic}}
\tikzset{sync/.style={color = sync}}
\tikzset{transition/.style={thick, color = black, align = center, font = \scriptsize},execute at begin node=\setlength{\baselineskip}{0.5ex}}
\tikzset{block/.style={rectangle, align = center, draw, fill=none, minimum width = 0.7cm, minimum height = 0.7cm, font=\footnotesize, thick} }
\tikzset{varblock/.style={rectangle, align = center, draw, fill=none, minimum width = 0.6cm, minimum height = 0.3cm, font=\bfseries\scriptsize, thick} }
\tikzset{simulinksignal/.style={thick,-latex}}
\tikzset{shatransition/.style={thick,-latex}}

\newcommand{\spacer}{0.6ex}

\input{config/misc}
\Crefname{figure}{Fig.}{Figs.}
\Crefname{section}{Sec.}{Secs.}
\usepackage{color}

\begin{document}
\title{Modeling Uncertainty: From Simulink to Stochastic Hybrid Automata\thanks{This research is partly funded by the DFG project  RealySt (471367371).}}
\titlerunning{Modeling Uncertainty in Simulink and SHA}
%
%
\author{Pauline Blohm\inst{1}~\Envelope~\orcidlink{0000-0001-8934-1861} \and
Felix Schulz\inst{1}  \and
Lisa Willemsen\inst{2} ~\orcidlink{0000-0002-0418-9854}   \and \\
Anne Remke\inst{1,2}~\orcidlink{0000-0002-5912-4767}\and
Paula Herber\inst{1,2}~\orcidlink{0000-0002-5349-154X}
}
\authorrunning{P. Blohm et al.}
%

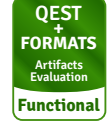
\institute{University of Münster, Germany, \\ 
\email{\{pauline.blohm, paula.herber, anne.remke\}@uni-muenster.de}\\
\email{felix.s.schulz@gmx.de}
\and University of Twente, The Netherlands, \email{l.c.willemsen@utwente.nl}}
\maketitle              
\begin{abstract}
Simulink is widely used in industrial design processes to model increasingly complex embedded control systems.
Thus, their formal analysis is highly desirable.
However, this comes with two major challenges: First, Simulink models often provide an idealized view of real-life systems and omit uncertainties such as, aging, sensor noise or failures.
Second, the semantics of Simulink is only informally defined.
In this paper, we present an approach 
to formally analyze safety and performance  of embedded control systems 
modeled in Simulink in the presence of uncertainty.
To achieve this, we  1) model different types of uncertainties as stochastic Simulink subsystems and 2) extend an existing formalization of the Simulink semantics based on stochastic hybrid automata (SHA) by providing transformation rules for the stochastic subsystems.
Our approach gives us access to established quantitative analysis techniques, like statistical model checking and reachability analysis.
We demonstrate the applicability of our approach by analyzing safety and performance in the presence of uncertainty for two smaller case studies.

    
\end{abstract}

\section{Introduction}
Embedded control systems require high functionality and flexibility, especially since they are  increasingly used in safety-critical environments, such as cars, airplanes, or energy control systems. Thus, formal verification is desirable to ensure their safety, performance and resilience.
Model-driven development tools such as MATLAB Simulink allow to graphically model and simulate complex hybrid control systems, i.e. systems that combine discrete and continuous behavior. 
One aspect that is often omitted when modeling real-life systems is their inherent uncertainty, e.g. caused by aging, sensor noise or failures.

Furthermore,  simulation executes the system  only for selected inputs and sound statistical methods like statistical model checking (SMC) are required to provide stochastic guarantees \cite{budde2024sound}.
However, existing approaches for reachability analysis and SMC of Simulink models either rely on a transformation of the Simulink model into a formal representation or perform SMC directly on the Simulink model. While the former approaches enable formal verification and even provide formal guarantees about crucial properties, they mostly disregard uncertainties and probabilistic behavior, thus, the results become useless in the presence of real-life effects such as aging, noise or failures. 
As Simulink does neither offer specification techniques for more complex path properties, nor hypothesis testing, it cannot be directly used for SMC. Existing approaches, e.g  \cite{zuliani2013bayesian}, apply SMC directly on the Simulink model, or include SMC methods in Simulink \cite{isola23}. However, statistical evaluations in Simulink are very costly due to the high overhead of simulations in Simulink. Furthermore, this does not provide a formal model which would be amenable to formal verification, like reachability analysis. 

In this paper, we present an approach to model uncertainties in Simulink and formalize them using stochastic hybrid automata (SHA). 
We build on previous work, where we proposed a modular and extensible transformation from Simulink to SHA, which gives us access to quantitative analysis methods. 
Previously, the transformation was only amenable for  simplified failure-repair models which illustrates the potential of introducing stochasticity into Simulink models.
Our contribution in this paper is twofold: 
1) We provide a library of Simulink subsystems for different types of stochastic behavior to capture uncertainties. 
2) We provide a formalization of the presented subsystems via a transformation into a dedicated SHA formalism that accommodates the stochastic extensions. Then, we seamlessly integrate the new transformation rules into our existing transformation from Simulink to SHA.
The transformation into SHA makes the whole approach amenable to both formal reachability analysis techniques 
and SMC. 
We demonstrate the feasibility of our approach by using the SMC tool \texttt{modes} \cite{Modes}  
on two small case studies, namely a temperature control system with sensor losses, and a simple energy measurement unit with stochastic switching. 

The rest of this paper is structured as follows: In \Cref{sec:bg}, we introduce the necessary background.  
In \Cref{sec:modeling}, we propose Simulink subsystems that are designed to model different types of uncertainties.
We present their respective formalization via SHA templates in \Cref{sec:formalization} and evaluate our approach in \Cref{sec:eval}. Finally, we summarize related work in \Cref{sec:relwork} and conclude in \Cref{sec:conclusion}.

\section{Background}
\label{sec:bg}
This section  introduces the necessary background for the remainder of this paper, namely Simulink and stochastic hybrid automata (SHA) and our transformation from Simulink to SHA.
\subsection{Simulink}
\begin{figure}[tb]
    \centering
    \begin{tikzpicture}

    \node[block,  minimum width = 1.2cm,minimum height = 1.2cm ] (switch) {};
    \draw[-, semithick] ($(switch.west)+(-0cm,0.4cm)$) to  ($(switch) + (-0.1cm,0.4cm)$);
    \draw[draw=black, semithick] ($(switch) + (-0.1cm,0.36cm)$) rectangle ++(0.08,0.08);
    \draw[-, semithick] ($(switch.west) + (0,-0.4cm)$) to ($(switch) + (-0.1cm,-0.4cm)$);
    \draw[draw=black, semithick] ($(switch) + (-0.1cm,-0.44cm)$) rectangle ++(0.08,0.08);    
    \draw[-, semithick] (switch) to node[right, font=\boldmath\scriptsize\bfseries]{$>$\,0} ($(switch.west) + (0.2cm,0)$);
    \draw[-, semithick] ($(switch) + (-0.02cm,0.4cm)$)  to ($(switch.east) + (-0.3cm,0)$)to (switch.east);

   \node[block, right = 2ex of switch, font=\boldmath] (int) {$\frac{1}{s}$};

   \node[block, left = 9ex of switch] (relay) {};
    \draw[-, color=lightgray!70, semithick] ($(relay.north) + (0,-0.04cm)$) to ($(relay.south) + (0,0.04cm)$);
    \draw[-, color=lightgray!70, semithick] ($(relay.west) + (0.04cm,0)$) to ($(relay.east) + (-0.04cm,0)$);
    \draw[-, semithick] ($(relay.west) + (0.04cm, -0.25cm)$) -| ($(relay) + (0.1cm, 0.25cm)$);
    \draw[-, semithick] ($(relay) + (-0.1cm, -0.25cm)$) |- ($(relay.east) + (-0.04cm, 0.25cm)$);    

    \node[block, left = 2ex of relay, font=\boldmath] (sum) {$+-$};

    \node[block,  left =2ex of sum] (tdes) {\scriptsize$21$\\ \tiny tdes};
    \node[block] (heat)at  ($(switch.west)+(-0.7cm,0.4cm)$) {\scriptsize0.03\\ \tiny heat}; 
    \node[block] (cool) at  ($(switch.west)+(-0.7cm,-0.4cm)$) {\scriptsize -0.03\\ \tiny cool};
    
    \draw[simulinksignal] (tdes.east) -- (sum.west);
     \draw[simulinksignal] (switch.east) --(int.west);
     \draw[simulinksignal] (heat.east) -- ([xshift=-0.3cm]switch.west |- heat.east) -- ([xshift=-0.3cm]$(switch.west)+(-0cm,0.4cm)$) --($(switch.west)+(-0cm,0.4cm)$);
     \draw[simulinksignal] (cool.east) -- ([xshift=-0.3cm]switch.west |- cool.east) -- ([xshift=-0.3cm]$(switch.west)+(-0cm,-0.4cm)$) --($(switch.west)+(-0cm,-0.4cm)$);
     \draw[simulinksignal] (sum.east) -- (relay.west);
    \ \draw[simulinksignal] (relay.east) -- (switch.west);
    \draw[simulinksignal] (int.east) -- ([xshift = 0.15cm] int.east) --([xshift = 0.15cm,yshift=-0.5cm] int.south east) --([yshift=-0.5cm] sum.south) -- (sum.south);
\end{tikzpicture}
    \caption{Simulink model of a temperature control system.}
    \label{fig:examplesimulink}
\end{figure}
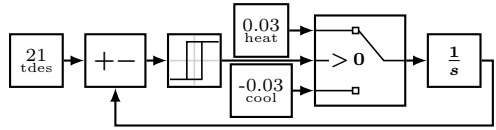
Simulink \cite{Simulink} is an industrially well established graphical modeling language for hybrid systems. It comes with a tool suite for simulation and automated code generation. 
Simulink models consist of blocks that are connected by discrete or continuous signals via ports. 
The Simulink block library provides a large set of predefined blocks, 
from arithmetics over control flow blocks to integrators and complex transformations.
Together with the MATLAB library, linear and non-linear differential equations
can be modeled and simulated.
Furthermore, the Simulink library provides random blocks to sample values from a probability distribution.
Simulink also provides the user with the option to define custom masks for subsystems, effectively allowing the user to create subsystems that can be parameterized and used  like regular Simulink blocks.

\paragraph{Example.}\Cref{fig:examplesimulink} shows a Simulink model of a temperature control system, which aims to keep the temperature in the room close to the desired temperature \emph{tdes}. Heating and cooling rates are modeled as constant blocks $heat$ and $cool$.
The system switches to heating if the temperature is below a specified lower threshold, and to cooling if it's above and upper threshold. A \emph{relay} block is used to prevent rapid switching, i.e.  the system only switches if the  temperature deviation is above a given tolerance. 

\subsection{Stochastic Hybrid Automata} 
SHA are an extension of hybrid automata (HA) with stochastic behavior. 
HA~\cite{alur1993hybrid} allow to capture the interaction of discrete and continuous behavior.  
Formally, they are defined in \cite{henzinger2000theory} as follows:
\begin{definition}[Hybrid Automata]
	\label{def:hybrid_automaton}
	A \emph{hybrid automaton (HA)} is a tuple $\mathcal{H} = (\Loc,\Var,\Flow,\allowbreak \Inv,\Lab,\Edge,\Init)$ with  components:
	\begin{itemize}
		\item $\Loc$ is a non-empty finite set of \emph{locations} or \emph{control modes}.
		\item $\Var=\{x_1,{\ldots},x_d \}$ is a finite ordered set
		of \emph{variables}. We call $\nu\,{\in}\,\mathcal{V}$ a \emph{valuation}, and $\state=(l,\nu)\in\Loc{\times}\mathcal{V} = \States$ a \emph{state} of $\H$. 
		\item $\Flow:\Loc\rightarrow (\mathcal{V} \rightarrow \mathcal{V})$ specifies for each location its \emph{flow} or \emph{dynamics}.
		\item $\Inv:\Loc\rightarrow 2^{\mathcal{V}}$ specifies an \emph{invariant} for each location. 
                \item $\Lab=\{a_1,\ldots,a_k\}$ is a non-empty finite ordered set of \emph{labels}.
		\item $\Edge \subseteq \Loc\times \Lab \times 2^{\mathcal{V}} \times (\mathcal{V}\rightarrow\mathcal{V})\times\Loc$ is
		a finite set of	\emph{edges}. For an edge $(l,a,g,r,l')\in\Edge$, $l$ and $l'$ are its
		\emph{source} resp. \emph{target}
		locations, $a$ its \emph{label}, $g$ its \emph{guard}, and $r$ its \emph{reset}.  Guards need to be disjoint for each pair of edges with identical source location and label.
		\item $\Init:\Loc\rightarrow 2^{\mathcal{V}}$ defines \emph{initial} valuations for each location. We call a state $(l,\nu)\in\States$ \emph{initial} if $\nu\in\Inv(l)\cap\Init(l)$.
	\end{itemize}
\end{definition}

Different formalisms exist to integrate stochastic behavior into HA. 
Here, we extend the definition of \emph{decomposed HA with eager non-predictive specification} (DHA) from~\cite{willemsen23Comparing,willemsen24camels} with a \emph{reset kernel} that allows to stochastically set the valuation of a continuous variable according to the current state of the automaton. 
For the required preliminaries from  probability theory we refer to \Cref{app:ProbabilityIntro}.

\begin{definition}[HA with stochastic kernels]
    A \emph{hybrid automaton  with \allowbreak stochastic kernels} (\hawk) is a tuple $\mathcal{A}=(\mathcal{H},\kernel,\resetKernel)$ with $\kernel=(\kernel_1,\ldots,\kernel_k)$, where:
	\begin{itemize}
	\item $\mathcal{H}=(\Loc, \Var, \Flow, \allowbreak\Inv, \Lab, \Edge, \Init)$ a HA with $ |\Var| =d$.
        \item $\kernel_i:\mathcal{B}(\Rpz)\times\States\to [0,1]$, $i=1,\ldots,k,$ where $k = |\Lab|$, are continuous stochastic kernels from $(\States,\allowbreak \mathcal{B}(\States))$ to $(\Rpz,\mathcal{B}(\Rpz))$, called \emph{delay} kernels.
        \item $\kernel^R:\mathcal{B}(\R^d)\times(\States\times\Lab)\to[0,1]$ is a continuous stochastic kernel from $((\States\times\Lab),\allowbreak \mathcal{B}((\States\times\Lab)))$ to $(\R^d,\mathcal{B}(\R^d))$, called \emph{reset} kernel.
	\end{itemize}
\end{definition}

The execution semantics of a \hawk follows the semantics of a DHA \cite{willemsen24camels}. Similarly to DHA, a \hawk $\mathcal{A}=(\H,(\kernel_1,\ldots,\kernel_k),\resetKernel)$ extends the underlying HA $\H$ with $k$ so-called \emph{random clocks} $c_1,\dots,c_k$. In each state, the $i$-th random clock evolves with  rate $1$ if an edge associated with label $a_i$ is enabled and with rate $0$ otherwise. The random clock $c_i$ is reset to $0$ if an edge associated with label $a_i$ is scheduled. 
During the execution of a \hawk,  the expiration time of the random clock $c_i$ is sampled based on the delay kernel $\kernel_i$ for the associated edge and  stored in a vector $\realisierungen$. 
An edge is taken if one random clock reaches its indicated expiration time, i.e.  if  $c_j = \realisierungen[j]$, for  $1\leq j\leq k$. 
For completeness, we provide a summary of the formal construction of DHA as defined in \cite{willemsen24camels} in \Cref{app:constrHEE}.
As an extension to DHA, \hawk include an additional \emph{reset step}, which directly follows each discrete step and immediately resets the continuous state of $\H$ according to the probability distribution given by the  reset kernel $\resetKernel$. 

In the following, we assume that the function  $\win(i)$ specifies  the index of the random clock which reached its indicated delay for the $i$-th step of the execution.
We denote the valuation of the continuous variables from $\H$ as $\sigma.\ass_{\H}$. 
For a \emph{probability density function (PDF)} $f : \Rpz \to \Rpz$ we define its \emph{support} as $\support(f) = \{\omega \in \R \suchthat f (\omega) > 0\}$. 
\begin{definition}[Semantics of \hawk]
  \label{def:decomposedEagerPredSemantic}
  A path $\pi$ of a given  \hawk $\mathcal{A}=(\H,(\kernel_1,\ldots,\kernel_k),\resetKernel)$, has the form $\pi= (\sigma_0,\realisierungen_0) \xrightarrow{t_0} (\sigma_0',\realisierungen_0)\xrightarrow{a_{\win(0)}} (\sigma_1',\realisierungen_1) \xrightarrow{r_0}(\sigma_1,\realisierungen_1) \xrightarrow{t_1}\ldots$ 
  such that 
         \begin{itemize}
         \item $\sigma_i\xrightarrow{t}\sigma_i'\xrightarrow{a_{\win(0)}}\sigma_{i+1}$ is governed by $\H$,
         \item $\realisierungen_i\in\Rpz^k$ for all $0\leq i\leq\dep(\pi)$,
         \item $\realisierungen_0[j]\in\support(\operatorname{Dist}^{\kernel_j}_{\sigma_{0}})$ for all $j\in\{1,\ldots,k\}$,
	 \item $\ass_i'(\auxwatch_{\win(i)})=\realisierungen_{i}[\win(i)]$, $\ass_i'(\auxwatch_{j})\leq \realisierungen_{i}[j]$, $\realisierungen_{i+1}[\win(i)]\in\support(\operatorname{Dist}^{\kernel_{\win(i)}}_{\sigma_{i+1}})$ and $\realisierungen_{i+1}[j]\allowbreak =\realisierungen_i[j]$ for all $0\leq i<\dep(\pi)$ and $j\in\{1,\ldots,k\}\setminus\{\win(i)\}$, 
     \item $\sigma_i.\ass_{\H}\in\support(\operatorname{Dist}^{\kernel^R}_{(\sigma_{i}\times\win(i))}) $ , and
         \item if $\pi$ is finite and it ends with a time step $(\sigma_i,\realisierungen_i) \xrightarrow{t_i} (\sigma_i',\realisierungen_i)$ then $\sigma_i'(\auxwatch_j)\leq \realisierungen_i[j]$ for all $j\in\{1,\ldots,k\}$.
	\end{itemize}
\end{definition}
To ease notation, we write $\resetKernel_{var_i}(\sigma,a)$ to indicate the probability measure specified by $\resetKernel$ for the continuous variable $var_i\in\Var $ in state $ \sigma \in \Sigma$ at edges with label $ a \in \Lab$. 

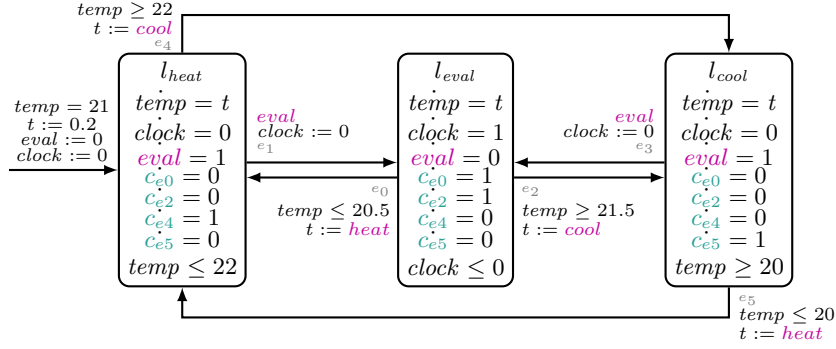
\begin{figure}[tb]
    \centering
     \begin{tikzpicture}
\useasboundingbox (-2.3,-2.3) rectangle  (8.9,2.3);
    \node[location] (l0) 
    {
    $l_{\mathit{heat}} $\\ [\spacer]
    $\dot{\mathit{temp}} = \mathit{t}$\\
    $\dot{\mathit{clock}} = 0$\\ 
    $\dot{\randomclock{eval}} = 1$\\ 
    $\dot{\randomclockH{c_{e0}}} = 0$\\ 
    $\dot{\randomclockH{c_{e2}}} = 0$\\ 
        $\dot{\randomclockH{c_{e4}}} = 1$\\ 
    $\dot{\randomclockH{c_{e5}}} = 0$\\         
    [\spacer]
    $\mathit{temp} \leq 22$
    };
    
    \node[location, right= 2cm  of l0] (l1) 
    {
    $l_{\mathit{eval}}$\\ [\spacer]
    $\dot{\mathit{temp}} = \mathit{t}$\\ 
    $\dot{\mathit{clock}} = 1$\\ 
    $\dot{\randomclock{eval}} = 0$\\
    $\dot{\randomclockH{c_{e0}}} = 1$\\ 
    $\dot{\randomclockH{c_{e2}}} = 1$\\ 
    $\dot{\randomclockH{c_{e4}}} = 0$\\ 
    $\dot{\randomclockH{c_{e5}}} = 0$\\         
    [\spacer]
    $\mathit{clock} \leq 0$
    };
    
     \node[location, right= 2cm  of l1] (l2) 
    {
    $l_{\mathit{cool}} $\\ [\spacer]
    $\dot{\mathit{temp}} = \mathit{t}$\\ 
    $\dot{\mathit{clock}} = 0$\\ 
    $\dot{\randomclock{eval}} = 1$\\
    $\dot{\randomclockH{c_{e0}}} = 0$\\ 
    $\dot{\randomclockH{c_{e2}}} = 0$\\ 
    $\dot{\randomclockH{c_{e4}}} = 0$\\ 
    $\dot{\randomclockH{c_{e5}}} = 1$\\      [\spacer]
    $\mathit{temp} \geq 20$
    };

    \draw[shatransition, sync] ([yshift =-0.1cm]l1.west) to node[below,transition,xshift =0.15cm, align = right] 
     {  {\tiny \color{gray} $e_0$}\\[0.5ex] $\mathit{temp} \leq 20.5$\\$\mathit{t} := \randomclock{heat}$  }  ([yshift = -0.1cm]l0.east);

     \draw[shatransition, sync] ([yshift = 0.1cm]l0.east) to node[above,transition, align = left, xshift = -0.25cm] 
     { $\randomclock{eval}$\\
     $\mathit{clock} := 0$\\[0.5ex]
     {\tiny \color{gray} $e_1$}} ([yshift = 0.1cm]l1.west);
    \draw[shatransition, sync] ([yshift = -0.1cm]l1.east) to node[below,transition, align = left, xshift = -0.15cm] 
     {   {\tiny \color{gray} $e_2$}\\[0.5ex]$\mathit{temp} \geq 21.5$\\ $\mathit{t} := \randomclock{cool}$  } ([yshift =-0.1cm]l2.west);

    \draw[shatransition, sync] ([yshift = 0.1cm]l2.west) to node[above,transition, align = right, xshift = 0.25cm] 
     { $\randomclock{eval}$\\
     $\mathit{clock} := 0$\\[0.5ex] 
     {\tiny \color{gray} $e_3$}} ([yshift = 0.1cm]l1.east) ;

    \draw[shatransition, sync, bend left] (l0.north) -- node[left,transition, yshift = 0.1cm, align = right] 
     {  $\mathit{temp} \geq 22$\\  $\mathit{t} := \randomclock{cool}$ \\[0.5ex] {\tiny \color{gray} $e_4$}  } ([yshift = 0.5cm] l0.north) -- ([yshift = 0.5cm] l2.north) -- (l2.north);
     
    \draw[shatransition, sync] (l2.south) -- node[right,transition,xshift = 0cm, yshift = -0.2cm, align = left] 
     {  {\tiny \color{gray} $e_5$}\\  $\mathit{temp} \leq 20$\\ $\mathit{t} := \randomclock{heat}$     }  ([yshift = -0.4cm]l2.south) -- ([yshift = -0.4cm]l0.south) --  (l0.south);

    \draw[shatransition, sync] ([xshift=-1.5cm]l0.west) to node[above,transition] 
     {  $\mathit{temp} =21$\\ $\mathit{t} := 0.2$  \\ $\mathit{eval} := 0$   \\ $\mathit{clock} := 0$  } (l0.west);
\end{tikzpicture}
    \caption{Simple Temperature Control Unit given as a \hawk.}
    \label{fig:sha-example}
\end{figure}

\paragraph{Example}
The \hawk shown in \Cref{fig:sha-example} models a simple temperature control unit that can either heat ($l_{\mathit{heat}}$) or cool ($l_{\mathit{cool}}$) the room with rate $\mathit{t}$ that depending on the state of the model is either chosen from $\mathcal{U}(0.1,0.3)$ or from $\mathcal{U}(-0.1,-0.3)$. 
If the temperature $\mathit{temp}$ is too high the system switches to cooling, and to heating if it gets to cold.
After a $\mathcal{N}(8,1)$-distributed delay, we move to $l_{\mathit{eval}}$ where the temperature is compared to given bounds. Then, if we move to $l_{\mathit{heat}}$ or $l_{\mathit{cool}}$, the rate $t$ is resampled according to the reset kernel. For each edge, we specify a delay kernel, which characterizes the distribution of the expiration time of the random clock associated with the edge.
Here, the delay kernels for the edges $e_i$ for $i \in \{0,2,4,5\}$ can be specified as 
$\kernel_{e_i}\sim \mathcal{D}(d_i)$ and for $i \in \{1,3\}$  as  $\kernel_{e_i}\sim \mathcal{N}(8,1)$.
For $i \in \{0,2\}$, i.e. for the urgent edges , $d_i = 0$.
For  $i \in \{4,5\}$, the delays $d_i$ are resolved such that $\int_0^{d_i} \mathit{t}\  dx + \nu(temp) = \theta_i$, where $\theta_i$ is the bound of the temperature at the discrete edge, i.e. $\theta_4 = 22$ for $e_4$ and $\theta_5 = 20$ for $e_5$ in our example.  
 The random clocks $\textit{eval}$, $c_{e0}$, $c_{e2}$, $c_{e4}$ and $c_{e5}$ track the enabling time of the corresponding edge.
Further, the reset kernels for each variable at each edge can be defined for all states $\sigma\in\States$ as follows: $\resetKernel_{\mathit{temp}}(\sigma,e_i) \sim\mathcal{D}(\mathit{temp})$, $ \resetKernel_{\mathit{clock}}(\sigma,e_i)\sim\mathcal{D}(\mathit{clock})$ for $i\in\{0,\dots,5\}$, $\resetKernel_{\mathit{t}}(\sigma,e_i) \sim \mathcal{D}(t)$ for  $i\in\{1,3\}$, $\resetKernel_{\mathit{t}}(\sigma,e_i)\sim \mathcal{U}(0.1,0.3)$ for  $i\in\{0,5\}$,  
and $\resetKernel_{\mathit{t}}(\sigma,e_i) \sim \mathcal{U}(-0.1,-0.3)$ for  $i\in\{2,4\}$.
To ease notation, we do not explicitly state the stochastic kernels that follow a Dirac distribution in the remainder of this paper, as their definition directly follows from the specification of the underlying HA, as illustrated in this example. 
Thus, we omit  (i) random clocks ($c_i$ for $i \in \{0,2,4,5\} $ in \Cref{fig:sha-example}), (ii) kernels specifying the stochastic delays  following a Dirac distribution and (iii) the definition of the stochastic reset kernel for Dirac distributed resets.

\subsection{Formalizing Simulink using SHA}
\label{sec:prelim-transformation}
To formally analyze Simulink models, we have previously proposed a modular transformation from Simulink to a subclass of SHA, namely \emph{linear hybrid automata with random clocks} (\LAC) \cite{blohm2025towardsQuantitative}. Note that \hawk are a conservative extension of \LAC with stochastic kernels, so each \LAC can easily be translated into a \hawk by specifying the corresponding delay kernel for each random clock.
The key idea of our transformation from Simulink to \LAC is as follows: The Simulink model is separated into the singular blocks and the signal flow.
Each block is transformed independently using transformation rules defined by so-called \emph{SHA templates}.
SHA templates are given as \LACsync, which extend \LAC by introducing synchronization labels as well as distinguishing between \emph{input} and \emph{output} variables and also relax the definition of the \LAC s' flow and initial state.
While output variables are used to model the signal driven by the corresponding block and thus have a known flow and initial value, input variables represent the signal lines connected to the inport. 
Therefore, they do not have a given flow or initial value.
As a result, the SHA templates do not have a defined execution semantics.
To maintain the execution order and correctly map the output variables to their corresponding input variables, a discrete-event synchronization via \emph{synchronization mappings} is derived from the signal lines.
Then, the SHA templates together with the synchronization mappings are composed using a modified parallel composition which results in a monolithic SHA following the \LAC formalism.
This automaton can then be analyzed with established tools for quantitative analysis, e.g. \realyst \cite{Delicaris2024} or \modest \cite{Modes}.

\section{Modeling Uncertainty in Simulink}
\label{sec:modeling}
Modeling real-life systems enables us to simulate component interaction and system behavior.
However, models often portray an idealized view of the real-life system and omit uncertainties like aging, sensor noise or failures.
To bridge this gap, 
we design Simulink subsystems that model different kinds of uncertainties using probability distributions. 
The subsystems are masked, i.e. they can be used in the same way as standard Simulink blocks, and different parameters and settings can be 
used to conveniently adjust them in a simple user interface. 

In \cite{blohm2025towardsIsola}, we have identified the following sources of uncertainties
in cyber-physical systems: 
measurement errors, noise, component failures,  
failed memory accesses, bit flips, clock errors and skews, and uncertain effects of chosen control values as well as uncertain physical effects. Conceptually, all of these uncertainties can be modeled via stochastic sampling of a signal or clock value. 
Measurement errors like noise can be modeled by adding a random value to a base signal. 
Failures can be modeled via a stochastic timeout after which a component fails. 
Bit flips and failed memory accesses can be modeled using stochastic switching between correct and failed bit or memory accesses. Clock errors or clock skews can be modeled using stochastic sampling, where sampling times are randomly chosen. Uncertain effects of control values as well as uncertain physical effects can be similarly modeled as noise and failures using stochastic noise or stochastic timeouts. 
Overall, to accommodate these uncertainties, we provide Simulink subsystems
for \emph{Stochastic Timers}, \emph{Stochastic Sampling}, \emph{Stochastic Noise}, and \emph{Stochastic Switching}. 
Furthermore, as many of these effects are worsening over time,
we provide subsystems for 
\emph{Discrete Aging} and \emph{Continuous Aging}. 
\footnote{The .slx-files of the subsystems are provided in the artifact.} 

\begin{figure}[tb]
    \centering
    \begin{subfigure}[t]{0.42\textwidth}
    \centering
    \begin{tikzpicture}    
   \node[block, font= \boldmath] (int) {$\frac{1}{s}$};
   \draw[->, semithick] ([xshift  = 0.05cm]int.west) -- ([xshift  = 0.2cm]int.west) -- ([xshift  = 0.2cm,yshift = -0.2cm]int.west);
   \draw[-, semithick] ([xshift  = 0.05cm]int.west) ([xshift  = 0.2cm,yshift = -0.2cm]int.west) |- ([yshift  = 0.1cm]int.south); 

    \node[block, right =2ex of int, font= \boldmath\scriptsize] (sum) {$-+$};

    \node[block,  font= \boldmath] (one) at ([xshift = -0.8cm,yshift = 0.2cm]int.west) {\scriptsize$1$};
    
    \node[block,  right =2ex of sum] (saturation) {};
    \draw[simulinksignal,-, color=lightgray!70,semithick] ($(saturation.north) + (0,-0.05cm)$) to ($(saturation.south) + (0,0.05cm)$);
    \draw[simulinksignal,-, color=lightgray!70,semithick] ($(saturation.west) + (0.05cm,0)$) to ($(saturation.east) + (-0.05cm,0)$);
    \draw[semithick,-] ($(saturation.west) + (0.1cm, -0.3cm)$) -- ($(saturation) + (-0.1cm, -0.3cm)$) -- ($(saturation) + (0.1cm, 0.3cm)$) -- ($(saturation.east) + (-0.1cm, 0.3cm)$) ;

    \node[port, right = 3ex of saturation] (outport) {$1$};
    \node[goto, below =1ex of one, font = \bfseries\boldmath\scriptsize, inner sep = 0.75mm] (random) {[\texttt{sample}]};

    \draw[simulinksignal] (saturation) -- (outport);
    \draw[simulinksignal] (sum) -- (saturation);
    \draw[simulinksignal] (int) -- (sum);
    \draw[simulinksignal] (random.east) -| (sum.south);
    
    \draw[simulinksignal]  ([xshift = 0.15cm]saturation.east) node[dot]{} --([xshift = 0.15cm,yshift = 0.2cm]saturation.north east)-- ([xshift = -0.3cm,yshift = 0.2cm]int.north west) --([xshift = -0.3cm,yshift = -0.2cm]int.west) --([yshift = -0.2cm]int.west) ;
    \draw[simulinksignal] (one.east) --([yshift = 0.2cm]int.west);
\end{tikzpicture}
    \caption{Stochastic Timer.}
    \label{fig:simulink-timer}
    \end{subfigure}%
    ~ 
    \begin{subfigure}[t]{0.58\textwidth}    
    \centering
    \begin{tikzpicture}    
    \node[subsystem,minimum width = 1.2cm, minimum height = 0.9cm,] (timer1) {};
    \node[blocklabel, below =0.1ex of timer1] (timer1L) {timer 1};
    \node[blocklabel, anchor = west] (trigger1) at (timer1.west) {};
    \node[blocklabel, anchor = east] (time1) at (timer1.east) {time};

    \node[subsystem,minimum width = 1.2cm, minimum height = 0.9cm,left = 0.25cm of timer1] (timer2) {};
    \node[blocklabel, below =0.1ex of timer2] (timer2L) {timer 2};
    \node[blocklabel, anchor = east] (trigger2) at (timer2.east) {time};
    \node[blocklabel, anchor = west] (time2) at (timer2.west) {};

    \node[block] (mem) at ([xshift = -1.2cm, yshift = 0cm]timer2) {};
    \draw[-latex, semithick] ([yshift=0.2cm]mem.south) -- ([xshift=-0.2cm, yshift=0.2cm]mem.south east) -- ([xshift=-0.2cm, yshift=-0.2cm]mem.north east) -- ([xshift=0.2cm, yshift=-0.2cm]mem.north west)--  ([xshift=0.2cm, yshift=0.2cm]mem.south west);

    \node[block,  minimum width = 1.2cm,minimum height = 1.2cm ,right =8ex of timer1] (switch) {};
    \draw[-, semithick] ($(switch.west)+(-0cm,0.4cm)$) to  ($(switch) + (-0.1cm,0.4cm)$);
    \draw[draw=black, semithick] ($(switch) + (-0.1cm,0.36cm)$) rectangle ++(0.08,0.08);
    \draw[-, semithick] ($(switch.west) + (0,-0.4cm)$) to ($(switch) + (-0.1cm,-0.4cm)$);
    \draw[draw=black, semithick] ($(switch) + (-0.1cm,-0.44cm)$) rectangle ++(0.08,0.08);    
    \draw[-, semithick] (switch) to node[right, font=\boldmath\scriptsize]{$ > 0$} ($(switch.west) + (0.2cm,0)$);
    \draw[-, semithick] ($(switch) + (-0.02cm,0.4cm)$)  to ($(switch.east) + (-0.3cm,0)$)to (switch.east);
    
    \node[port, anchor = east] (inport1) at ($(switch.west)+(-0.25cm,0.4cm)$)  {$1$};
    \node[port, anchor = east] (inport2) at ($(switch.west) + (-0.25cm,-0.4cm)$)  {$2$};
    \node[port, right = 1.5ex of switch] (outport) {$1$};

    \draw[simulinksignal] (inport1) -- ($(switch.west)+(0cm,0.4cm)$);
    \draw[simulinksignal] (inport2) -- ($(switch.west) + (0cm,-0.4cm)$) ;    
    \draw[simulinksignal] (switch) -- (outport);
     
    \draw[simulinksignal] (timer1.east) --(switch);
    \draw[simulinksignal] (mem.east) -- (timer2);
    \draw[simulinksignal] (timer2) -- (timer1);
    \draw[simulinksignal] (timer1.east) -| node[dot]{} ([xshift = 0.15cm, yshift = -0.3cm]timer1.south east)-|(mem.south);
\end{tikzpicture}
    \caption{Stochastic Switching.}
    \label{fig:simulink-switching}
    \end{subfigure}
    \caption{Simulink Subsystems for Timer and Switching.}
\end{figure}
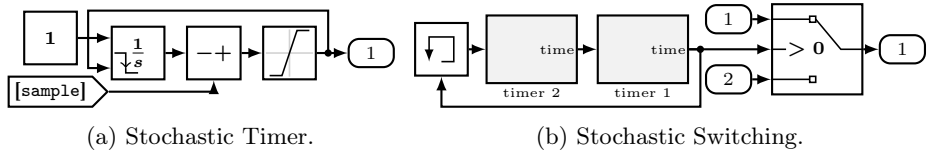

\paragraph{Stochastic Timer.} 
The subsystem shown in \Cref{fig:simulink-timer} models a timer whose expiration time is sampled from a probability distribution.
The stochastic timer is used in most of the stochastic subsystems presented in the following to model a randomly distributed time delay. 
It functions as follows: Upon a trigger, an expiration time is sampled from a  probability distribution and an integrator block is reset to zero.
The integrator block functions as a clock by integrating over the value 1. 
Subtracting the clock value from the expiration time gives the current value of the timer, provided as the output. 
Once the timer reaches zero, the falling edge re-triggers this subsystem.
The stochastic timer supports selecting between a uniform distribution, with configurable minimum and maximum parameters, or a folded-normal distribution with $\mathit{mean}= 0$ and a configurable variance. 
The seed for the random number generators can be set in the mask.

\paragraph{Stochastic Switching.}
The subsystem shown in \Cref{fig:simulink-switching} models a stochastic extension of the switch block provided by Simulink.
In contrast to the regular switch block, the switching condition for the stochastic switch does not depend on a third signal but rather on stochastic delays provided by two stochastic timers.
While the value of the first timer is greater than zero, i.e. it has not yet expired, the signal provided at the first inport is passed to the outport. 
Similarly, while the value of the second timer is greater than zero, the signal provided at the second inport is passed to the outport.
Once a timer expires, an expiration time for the other timer is sampled according to its distribution, which is specified in the mask.
The stochastic switch will initially sample an expiration time for the first timer and therefore output the signal at the first inport.
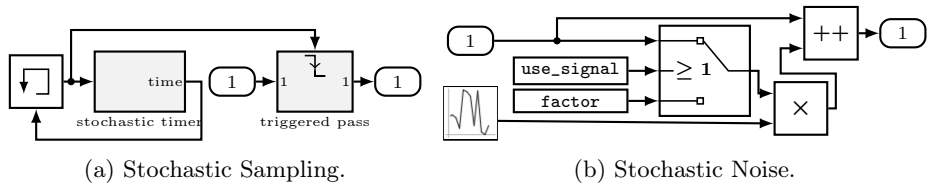
\begin{figure}[b]
    \centering
    \begin{subfigure}[t]{0.45\textwidth}    
    \centering
    \begin{tikzpicture}    
    \node[subsystem,minimum width = 1.2cm, minimum height = 0.8cm,] (timer) {};
    \node[blocklabel, below =0.1ex of timer] (timerL) {stochastic timer};
    \node[blocklabel, anchor = west] (trigger) at (timer.west) {};
    \node[blocklabel, anchor = east] (time) at (timer.east) {time};

    \node[port, ,right = 0.3cm of timer] (inport)  {$1$};
    \node[subsystem,minimum width = 1cm, minimum height = 0.8cm,right = 2ex of inport] (delay) {};
    \node[blocklabel, below =0.1ex of delay] (delayL) {triggered pass};
    \node[blocklabel, anchor = west] (in) at (delay.west) {1};
    \node[blocklabel, anchor = east] (out) at (delay.east) {1};
    \draw[->, semithick] ([xshift = -1ex,yshift = -0.5ex]delay.north) -| ([yshift = -1.8ex]delay.north);
    \draw[-, semithick] ([yshift = -1.8ex]delay.north) |- ([xshift = 1ex,yshift = -2.5ex]delay.north) -| ([yshift = -2ex]delay.north);
    \node[port, right = 2ex of delay] (outport) {$1$};

    \node[block, left = 0.4cm of timer] (mem) {};
    \draw[-latex, semithick] ([yshift=0.2cm]mem.south) -- ([xshift=-0.2cm, yshift=0.2cm]mem.south east) -- ([xshift=-0.2cm, yshift=-0.2cm]mem.north east) -- ([xshift=0.2cm, yshift=-0.2cm]mem.north west)--  ([xshift=0.2cm, yshift=0.2cm]mem.south west);delay.north

    \draw[simulinksignal] (inport) -- (delay);
    \draw[simulinksignal] (delay) -- (outport);
    \draw[simulinksignal] (mem) -- (timer);
    \draw[simulinksignal] ([xshift=0.1cm]mem.east) node[dot]{} |- ([yshift=2ex]delay.north) -- (delay.north);
    \draw[simulinksignal] (timer.east) -- ([xshift = 1.5ex]timer.east) |- ([yshift = -3ex]mem.south)-- (mem.south);
\end{tikzpicture}
    \caption{Stochastic Sampling.}
    \label{fig:simulink-delay}
    \end{subfigure}%
    ~ 
    \begin{subfigure}[t]{0.55\textwidth}    
    \centering
    \includestandalone[mode=image|tex] {tikz/simulink_models/simulink_noise}
    \caption{Stochastic Noise.}
    \label{fig:simulink-noise}
    \end{subfigure}
    \caption{Simulink Subsystems for Sampling and Noise.}
\end{figure}
\paragraph{Stochastic Sampling.}
This subsystem shown in \Cref{fig:simulink-delay} models a random sampling of a signal.
A stochastic timer is used to provide randomly distributed sampling times.
When the timer expires, the value of the output signal is updated to the current value of the input signal and the timer is triggered to sample a new expiration time.
While the timer is running, the output signal holds the latest value.
Exemplary, this system can be used to model a sensor with a non-constant sampling rate.
The mask for this subsystems allows the user to configure the stochastic timer.
\paragraph{Stochastic Noise.}
The subsystem shown in \Cref{fig:simulink-noise} models a signal distorted by noise  sampled from a normal distribution. 
This subsystem uses Simulinks' random number generator to sample random numbers from a normal distribution at specified discrete intervals.
The random number is then either multiplied with a constant noise factor or with the input signal to calculate the amount of distortion which is then added to the input signal.
The subsystems mask allows the designer to enable and disable the noise, specify the noise factor, mean and variance for the distribution as well as the sample time and a seed.
\begin{figure}[tb]
    \centering
\begin{subfigure}[t]{1\textwidth}
    \centering
    \begin{tikzpicture}    
    \node[subsystem,minimum width = 1.5cm, minimum height = 0.9cm,] (timer1) {};
    \node[blocklabel, below =0.1ex of timer1] (timer1L) {stochastic timer};
    \node[blocklabel, anchor = west] (trigger1) at (timer1.west) {trigger};
    \node[blocklabel, anchor = east] (time1) at ([yshift = 0.2cm]timer1.east) {failed};
    \node[blocklabel, anchor = east] (time2) at ([yshift = -0.2cm]timer1.east) {count};

    \node[subsystem,minimum width = 1.5cm, minimum height = 0.9cm,below = 0.5cm of timer1] (timer2) {};
    \node[blocklabel, below =0.1ex of timer2] (timer2L) {stochastic timer};
    \node[blocklabel, anchor = east] (trigger2) at (timer2.east) {trigger};
    \node[blocklabel, anchor = west] (time2) at (timer2.west) {repair};

    \node[block] (mem) at ([xshift = -1.5cm, yshift = -0.7cm]timer1) {};
    \draw[-latex, semithick] ([yshift=0.2cm]mem.south) -- ([xshift=-0.2cm, yshift=0.2cm]mem.south east) -- ([xshift=-0.2cm, yshift=-0.2cm]mem.north east) -- ([xshift=0.2cm, yshift=-0.2cm]mem.north west)--  ([xshift=0.2cm, yshift=0.2cm]mem.south west);

    \node[subsystem,minimum width = 1.2cm, minimum height = 0.7cm] (age) at ([xshift = 22ex, yshift = -0.3cm] mem.east){};
    \node[blocklabel, below =0.1ex of age] (ageL) {age factor};
    \node[blocklabel, anchor = east] (factor) at (age.east) {factor};


    \node[block, font= \boldmath] (prod2) at ([xshift = 0.7cm, yshift = 0.2cm]age.east) {$\times$};
    \node[block, font= \boldmath] (sum2) at([xshift =16ex, yshift = 0.4cm] prod2) {$+-$};
    \node[block,  font= \boldmath] (one2) at ([xshift =-1cm,yshift = 0.2cm] sum2.west) {\scriptsize$1$};
    \node[block,font= \boldmath] (prod3) at ($(sum2.east)+(6ex,-0.2cm)$)  
    {$\times$};

    \node[block, minimum width = 1.2cm,minimum height = 1.3cm] (switch) at ([xshift = 7ex,yshift=-0.5cm ]prod3.east) {};
    \draw[-, semithick] ($(switch.west)+(-0cm,0.5cm)$) to  ($(switch) + (-0.1cm,0.5cm)$);
    \draw[draw=black, semithick] ($(switch) + (-0.1cm,0.46cm)$) rectangle ++(0.08,0.08);
    \draw[-, semithick] ($(switch.west) + (0,-0.5cm)$) to ($(switch) + (-0.1cm,-0.5cm)$);
    \draw[draw=black, semithick] ($(switch) + (-0.1cm,-0.54cm)$) rectangle ++(0.08,0.08);    
    \draw[-, semithick] (switch) to node[right, font=\boldmath\scriptsize]{$ > 0$} ($(switch.west) + (0.2cm,0)$);
    \draw[-, semithick] ($(switch) + (-0.02cm,0.5cm)$)  to ($(switch.east) + (-0.3cm,0)$)to (switch.east);
    
    \node[port, right = 2ex of switch] (outport) {$1$};
    \node[subsystem,minimum width = 1.5cm, minimum height = 0.8cm, font = \scriptsize] (repair) at ($(switch.west)+ (-1.2,-0.4cm)$) {};
    \node[blocklabel, anchor = west] (repairIn) at ([yshift = 0.2cm]repair.west) {input};
    \node[blocklabel, anchor = west] (repairExt) at ([yshift = -0.2cm]repair.west) {external};
    \node[blocklabel, below =0.1ex of repair] (repairL) {choose repair};
    
    \node[port,minimum height = 0.2cm] (inport1) at ([xshift = -2cm, yshift = 0.25cm]repair){$1$};
    
    \node[port,minimum height = 0.2cm] (inport2) at ([xshift = -2cm, yshift = -0.25cm]repair) {$2$};

    \draw[simulinksignal] (switch) -- (outport);
     
    \draw[simulinksignal] ([yshift = 0.2cm]timer1.east) --([xshift = 0.3cm,yshift = 0.2cm]timer1.east) |- (timer2.east);
    \draw[simulinksignal] (timer2.west) -| (mem.south);
    \draw[simulinksignal] (mem.north) |- (timer1.west);
   
    \draw[simulinksignal] (prod3) -- ($(switch.west)+(0cm,0.5cm)$);
    \draw[simulinksignal] ([yshift = -0.2cm]timer1.east) -| ([xshift = -1.5ex,yshift = 0.2cm]prod2.west) -- ([yshift = 0.2cm]prod2.west);
    \draw[simulinksignal] (sum2.east) -- ([yshift = 0.2cm]prod3.west);
    \draw[simulinksignal] (inport1) --([ yshift = 0.25cm]repair.west);
    \draw[simulinksignal] (inport2) -- ([yshift = -0.25cm]repair.west);
    \draw[simulinksignal] (repair.east) -- ([xshift = 1ex]repair.east)|- ($(switch.west)+(0cm,-0.5cm)$);
    \draw[simulinksignal] (one2.east) -- ([xshift = 1ex]one2.east)|- ([ yshift = 0.2cm]sum2.west);
    \draw[simulinksignal] (prod2.east) -- ([xshift = 6ex]prod2.east)|- ([ yshift = -0.2cm]sum2.west);
    \draw[simulinksignal] ([xshift = 1ex] inport1.east) node[dot]{} -- ([xshift = 1ex,yshift = 2ex] inport1.east)-|  ([xshift = -2ex,yshift = -0.2cm]prod3.west)--([yshift = -0.2cm]prod3.west);
    \draw[simulinksignal] ([xshift = 0.3cm,yshift = 0.2cm]timer1.east) node[dot] {} -| ([xshift = -2ex]switch.west)-- (switch.west);
    \draw[simulinksignal] (age) -- ([yshift = -0.2cm]prod2.west);
    \end{tikzpicture}
    \caption{Stochastic Discrete Aging.}
    \label{fig:simulink-agingdiscrete}
\end{subfigure}\\
\begin{subfigure}[t]{1\textwidth}
    \centering
    \begin{tikzpicture}    
    \node[subsystem,minimum width = 1.5cm, minimum height = 0.7cm] (integrator) {};
    \node[blocklabel, below =0.1ex of integrator] (integratorL) {age integrator};
    \node[blocklabel, anchor = east] (age) at (integrator.east) {age};
    \draw[-] ([xshift = -0.12cm,yshift = -0.15cm]integrator.north) -| ([xshift = -0.05cm,yshift = -0.07cm]integrator.north) --([xshift = 0.05cm,yshift = -0.07cm]integrator.north) |- ([xshift = 0.12cm,yshift = -0.15cm]integrator.north);
    
    \node[subsystem,minimum width = 1.5cm, minimum height = 0.7cm,] (timer) at ([xshift = 2.3cm]integrator) {};
    \node[blocklabel, below =0.1ex of timer] (timerL) {controll timer};
    \node[blocklabel, anchor = west] (trigger) at (timer.west) {age};
    \node[blocklabel, anchor = east] (time1) at ([yshift = 0.2cm]timer.east) {aging};
    \node[blocklabel, anchor = east] (repairtrigger) at ([yshift = -0.2cm]timer.east) {trigger};

    \node[subsystem,minimum width = 1.2cm, minimum height = 0.7cm] (agepause) at ([yshift =0.7cm, xshift = -0.8cm]integrator.north) {};
    \node[blocklabel, anchor = east] (pause) at (agepause.east) {pause?};
    \node[blocklabel, below =0.1ex of agepause] (agepauseL) {pause age};    
    \draw[-] ([xshift = -0.12cm,yshift = -0.15cm]agepause.north) -| ([xshift = -0.05cm,yshift = -0.07cm]agepause.north) --([xshift = 0.05cm,yshift = -0.07cm]agepause.north) |- ([xshift = 0.12cm,yshift = -0.15cm]agepause.north);
    
    \node[subsystem,minimum width = 1.6cm, minimum height = 0.7cm] (stochastictimer) at ([xshift = 1.7cm]repairtrigger) {};
    \node[blocklabel, below =0.1ex of stochastictimer] (stochastictimerL) {stochastic timer};
    \node[blocklabel, anchor = east] (stochastictimerrepair) at (stochastictimer.east) {repair};
    \node[blocklabel, anchor = west] (stochastictimertoggle) at (stochastictimer.west) {};

    \node[block] (mem) at ([xshift = 1cm] stochastictimerrepair) {};
    \draw[-latex, semithick] ([yshift=0.2cm]mem.south) -- ([xshift=-0.2cm, yshift=0.2cm]mem.south east) -- ([xshift=-0.2cm, yshift=-0.2cm]mem.north east) -- ([xshift=0.2cm, yshift=-0.2cm]mem.north west)--  ([xshift=0.2cm, yshift=0.2cm]mem.south west);
    
    \node[block, minimum width = 1.2cm,minimum height = 1.3cm] (switch) at (8.6cm,0.8cm) {};
    \draw[-, semithick] ($(switch.west)+(-0cm,0.5cm)$) node[] (switchIn1) {} to  ($(switch) + (-0.1cm,0.5cm)$);
    \draw[draw=black, semithick] ($(switch) + (-0.1cm,0.46cm)$) rectangle ++(0.08,0.08);
    \draw[-, semithick] ($(switch.west) + (0,-0.5cm)$) node[] (switchIn3) {} to ($(switch) + (-0.1cm,-0.5cm)$);
    \draw[draw=black, semithick] ($(switch) + (-0.1cm,-0.54cm)$) rectangle ++(0.08,0.08);    
    \draw[-, semithick] (switch) to node[right, font=\boldmath\scriptsize]{$ > 0$} ($(switch.west) + (0.2cm,0)$);
    \draw[-, semithick] ($(switch) + (-0.02cm,0.5cm)$)   to ($(switch.east) + (-0.3cm,0)$)to (switch.east);
    
    \node[subsystem,minimum width = 1.5cm, minimum height = 0.7cm, font = \scriptsize] (chooseRepair) at ([xshift = -1.4cm]switchIn1) {};
    \node[blocklabel, anchor = west] (chooseRepairIn) at ([yshift = 0.2cm]chooseRepair.west) {input};
    \node[blocklabel, anchor = west] (chooseRepairExt) at ([yshift = -0.2cm]chooseRepair.west) {external};
    \node[blocklabel, below =0.1ex of chooseRepair] (rchooseRepairL) {choose repair};
    
    \node[port, right = 2ex of switch] (outport) {$1$};    
    \node[port, left = 16ex of chooseRepairIn] (signal) {$1$};
    \node[port, left = 2ex of chooseRepairExt] (repairsignal) {$2$};

    \node[block] (saturation) at ([xshift = -0cm,yshift = -0.7cm]signal) {};
    \draw[simulinksignal,-, color=lightgray!70] ($(saturation.north) + (0,-0.05cm)$) to ($(saturation.south) + (0,0.05cm)$);
    \draw[simulinksignal,-, color=lightgray!70] ($(saturation.west) + (0.05cm,0)$) to ($(saturation.east) + (-0.05cm,0)$);
    \draw[simulinksignal,-] ($(saturation.west) + (0.1cm, -0.3cm)$) -- ($(saturation) + (-0.1cm, -0.3cm)$) -- ($(saturation) + (0.1cm, 0.3cm)$) -- ($(saturation.east) + (-0.1cm, 0.3cm)$) ;
    
    \node[block, font= \boldmath] (prod) at ([xshift = 1.2cm, yshift = -0cm]saturation) {$\times$};
    
    \draw[simulinksignal] (repairtrigger) -- (stochastictimer);
    \draw[simulinksignal] (age) -- (timer);
    \draw[simulinksignal] (stochastictimerrepair) -- (mem);
    \draw[simulinksignal] (mem.east) -| ([xshift = 1.3cm, yshift = -.3cm]mem.south east) -| ([xshift = -0.3cm]integrator.west) --(integrator.west);
    \draw[simulinksignal]  ([xshift = 1.3cm]mem.east) node[dot]{}|-(switch);
    
    \draw[simulinksignal] (chooseRepair) -- (switchIn1.center);
    \draw[simulinksignal] (switch) -- (outport);
    \draw[simulinksignal] (signal) -- (chooseRepairIn);
    \draw[simulinksignal] (repairsignal) -- (chooseRepairExt);
    \draw[simulinksignal] (prod.east) -- ([xshift = 2ex] prod.east) |-(switchIn3.center);
    \draw[simulinksignal] ([xshift = 2ex]signal.east) node[dot]{} |- ([yshift = 0.2cm]prod.west);
    \draw[simulinksignal] ([xshift =2ex]integrator.east) node[dot]{}|- (saturation.west);
    \draw[simulinksignal] (saturation.east) -- ([xshift = 1.5ex]saturation.east) |-([yshift = -0.2cm]prod.west);

    \draw[simulinksignal] (pause) -- ([xshift = 2ex]pause)-| (integrator.north);

    \draw[simulinksignal] (time1.east) -| ([xshift = 0.15cm, yshift = -0.35cm]timer.south east)-| ([xshift=-0.15cm, yshift = 0.25cm]agepause.north west) -| (agepause.north);
    \end{tikzpicture}
    \caption{Stochastic Continuous Aging}
    \label{fig:simulink-agingcontinuous}
\end{subfigure}
    \caption{Simulink Subystems for Discrete and Continuous Aging.}
\end{figure}
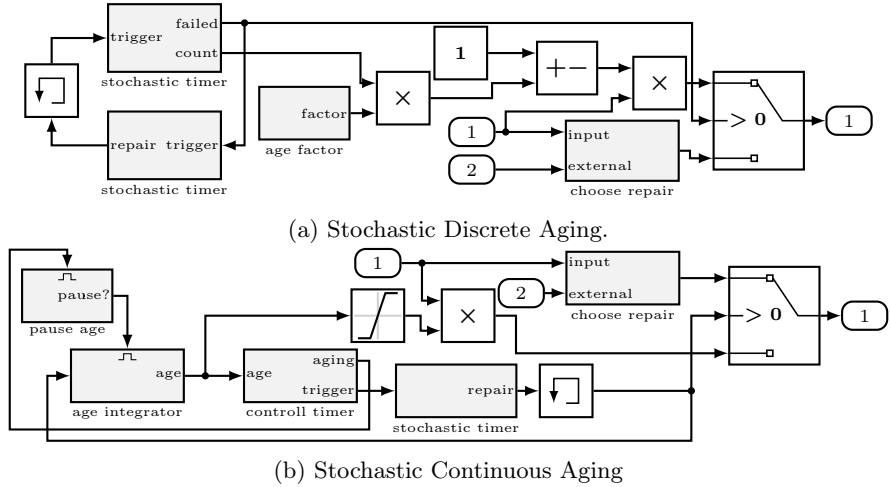
\paragraph{Stochastic Discrete Aging.}
The subsystem shown in \Cref{fig:simulink-agingdiscrete} models a signal that is degraded by repeatedly reducing the  input signal by an aging factor until a lower bound is reached that triggers a repair of the signal. 
The aging factor is increased  every time a stochastic timer expires.
The lower bound is derived from the maximum number of times the factor can be decreased. 
Once this bound is reached, a repair is triggered whose duration is provided by a stochastic timer.
During the repair, either an explicit repair signal is passed to the outport or a specified percentage of the input signal.
The mask allows the designer to specify a multitude of options such as the underlying distributions for the reduction time steps and the repair time, the maximum number of reduction steps or whether an explicit repair signal is used.
\paragraph{Stochastic Continuous Aging.}
The subsystem shown in \Cref{fig:simulink-agingcontinuous} also models signal degradation up to a specified lower bound.
However, in this case the signal is degraded linearly based on the time passed since the last repair.
Similarly to the discrete aging, once a lower bound is reached a repair of the signal is triggered and during the repair the designer can choose whether an explicit repair signal or a specified percentage of the input signal is passed to the outport.
Additionally, the signal degradation can also be paused and resumed based on a stochastic timer.
The mask enables the designer to customize 
the distributions of the different timers, whether an external repair signal is used and the rate of degradation.
\section{Formalizing Stochastic Subsystems using SHA}
\label{sec:formalization}
To formalize the presented Simulink subsystems, we use SHA and build upon an existing modular transformation from Simulink to SHA presented in \Cref{sec:prelim-transformation}.
In particular, we provide individual transformation rules that translate each of the Simulink subsystems into a SHA template. 
To enable seamless integration of these transformation rules into our existing transformation, we first lift 
the parallel composition of SHA templates from \LACsync to \hawksync to account for the extended stochastic behavior.
In the following, we first explain this lifting and then the individual transformation rules for each Simulink subsystem.

\subsection{Lifting SHA Template Composition to \hawk} 
\label{sec:extension}
While the concept of our transformation from Simulink to SHA is generally applicable to a wide range of automata classes, some adjustments are necessary to correctly compose the more complex stochastic behavior of \hawk.
First, to define SHA templates 
we introduce \hawksync  analogously to the \LACsync used in \cite{blohm2025towardsQuantitative}.
Intuitively, \hawksync extend the definition of \hawk by introducing synchronization labels to indicate sending and receiving of variables and splitting the variable set into input and output variables, and relax it by only requiring flow and initial value for output variables.
For \hawksync, the delay and reset kernels are also only defined for output variables.
A formal definition is provided in \Cref{app:transformation}, \Cref{def:hawksync}. 
Note, that analogous to \LACsync, \hawksync also do not have an execution semantics.

To correctly compose the \hawksync using the previously proposed transformation, the rules for synchronized and non-synchronized edges presented in \cite{blohm2025towardsQuantitative} need to be lifted to include the stochastic kernels.
Intuitively, the rules are extended as follows:
For non-synchronized edges we maintain the delay and reset kernels defined for this edge in the SHA templates.
For synchronized edges, we assign the continuous kernel from the sending edge for the resulting synchronized edge.
When assigning the reset kernel to the new edge we combine all reset kernels from the individual edges.
This is possible as the reset kernel of a \hawksync only considers the output variables which are unique for each SHA template.
For a more formal definition please see \Cref{app:transformation} \Cref{def:nonsyncjump} and \Cref{def:syncjump}.
The result of the composition is a monolithic SHA which follows the definition of a \hawk.

\subsection{Individual Transformation Rules for each Stochastic Subsystem}
We present  SHA templates that formalize the semantics of the stochastic Simulink subsystem presented in \Cref{sec:modeling}. The graphical illustrations indicate random clocks or sampling from non-Dirac distributions in pink, receiving of the corresponding variable in light blue and sending of the corresponding variable in green. Parameters provided by the Simulink block are \simulink{represented} accordingly. If not stated otherwise, $\Init(l) = \{ \emptyset, \texttt{false}\} $ for any location $l$.

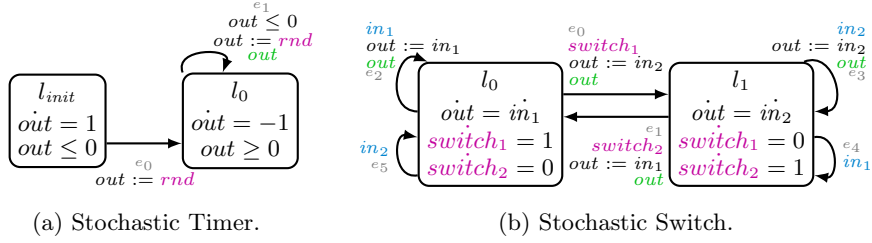
\begin{figure}[tb]
    \centering
    \begin{subfigure}[t]{0.3\textwidth}
    \centering
    \begin{tikzpicture}
    \node[location] (l0) 
    {
    $l_0$\\ [\spacer]
    $\dot{out} = -1$\\ [\spacer]
    $out \geq 0$
    };
    
    \node[location, left= 1cm  of l0] (l1) 
    {
    $l_{\mathit{init}}$\\ [\spacer]
    $\dot{out} = 1$\\ [\spacer]
    $out \leq 0$
    };

    \draw[shatransition, sync, bend left = 110, looseness = 1.7] (l0.140) to node[right,transition,xshift =0cm,yshift=0.3cm] 
    {{\tiny \color{gray}$e_1$}\\$\out \leq 0$\\
    $\out := \randomclock{rnd} $ \\
    $\send{\out}$} (l0.110);

    \draw[shatransition, sync] ([yshift=-0.3cm]l1.east) to node[below,transition, yshift=-0.1cm] 
     {    {\tiny \color{gray}$e_0$}\\$\out := \randomclock{rnd}$     }  ([yshift=-0.3cm]l0.west);
    
\end{tikzpicture}
    \caption{Stochastic Timer.}
    \label{fig:sha-timer}
    \end{subfigure}%
    ~ 
    \begin{subfigure}[t]{0.7\textwidth}    
  
    \centering
    \begin{tikzpicture}
    \node[location] (l0) {$l_0$ \\ [\spacer] $\innervar{\dot{\out}}  =\outervar{\dot{\inone}}$ \\ $\randomclock{\dot{switch_1}} = 1 $  \\ $\randomclock{\dot{switch_2}} = 0 $ };
    \node[location, right=1.4cm  of l0] (l1) {$l_1$ \\ [\spacer] $\innervar{\dot{\out}}  =\outervar{\dot{\intwo}}$ \\ $\randomclock{\dot{switch_1}} =0 $ \\ $\randomclock{\dot{switch_2}} = 1 $ };

    \draw[shatransition,sync,] ([yshift=0.4cm]l0.east) to node[above, transition, align = left] {{\tiny\color{gray} $e_0$}\\$\randomclock{switch_1}$\\  $\innervar{\out} := \outervar{\intwo} $\\ $\send{\out}$}  ([yshift=0.4cm]l1.west);
    
    \draw[shatransition, sync,] ([yshift=0.1cm]l1.west) to node[below, transition, align = right] {{\tiny\color{gray} $e_1$}\\$\randomclock{switch_2}$ \\  $\innervar{\out} := \outervar{\inone} $\\ $\send{\out}$}  ([yshift=0.1cm]l0.east);
    
    \draw[shatransition, sync, bend left = 100, looseness = 1.7] (l0.170) 
    to node[above,transition,xshift =0.2cm,yshift=-0.1cm, align = left] {    
    $\receive{\inone} $ \\
    $\innervar{\out} := \outervar{\inone} $\\ 
    $\send{\out}$\\
    {\tiny\color{gray} $e_2$}
    } 
    (l0.135);
    
    \draw[shatransition,sync, bend left = 100, looseness = 1.7] (l1.45) to node[above, transition, xshift= -0.2cm,yshift=-0.1cm, align = right] {
    $\receive{\intwo} $ \\
    $\innervar{\out} := \outervar{\intwo} $\\ 
    $\send{\out}$\\
    {\tiny\color{gray} $e_3$}}  (l1.10);
    
    \draw[shatransition, sync, bend left =100, looseness = 1.7] (l1.-10) to node[right,transition,xshift = 0cm,yshift=0cm, align = left] {{\tiny\color{gray} $e_4$}\\$\receive{\inone}$ }  (l1.-35);
    
    \draw[shatransition, sync, bend left = 100, looseness = 1.7] (l0.215) 
    to node[left,transition,xshift = 0cm,yshift=0.cm, align =right] 
    {
    $\receive{\intwo}$ \\[0.5mm]
    {\tiny\color{gray} $e_5$}
    }  
    (l0.190);
\end{tikzpicture}
    \caption{Stochastic Switch.
    }
    \label{fig:sha-switch}
    \end{subfigure}
    \caption{SHA Templates for Timer and Switch.}
\end{figure}
\paragraph{Stochastic Timer.}
To formalize the stochastic timer subsystem (see \Cref{fig:simulink-timer}), we define the SHA template shown in \Cref{fig:sha-timer}.
The variable \out represents the value of the output signal which is initially sampled from the distribution specified in the mask of the subsystem. 
To realize this initial sampling, the urgent location $l_{\mathit{init}}$ is added where no time can pass and the immediate edge to $l_0$ assigns the value of \out.
This value  then decreases with a rate of -1 until it reaches zero, which results in taking the self-loop  where a new random value is assigned to \out  and the discrete update of \out is sent.
As the Simulink block does not have any input, the template also does not have any input variables.
The reset kernel is defined as $\Psi^R_{\out} (\sigma,e_i)\sim \simulink{Dist} $ for $i \in \{0,1\}$ where $ \simulink{Dist}=\mathcal{U}(\simulink{low}, \simulink{high}) \mid \mathcal{N}_{\geq 0} ~(\simulink{var}) $ for all states $\sigma$ and  $\Init(l_{\mathit{init}}) = (\{\out= 0\}, \texttt{true})$.

\paragraph{Stochastic Switch.}
To formalize the stochastic switch subsystem (see \Cref{fig:simulink-switching}), we define the SHA template shown in \Cref{fig:sha-switch}.
Each location represents one of the two cases for the switch, i.e. either the variable \out has the same value as \inone or as \intwo.
Switching between these locations depends on the random clocks \randomclock{$\mathit{switch1}$} and \randomclock{$\mathit{switch2}$}. 
Initially, the value of \inone is assigned to \out.
Once the expiration time for \randomclock{$\mathit{switch1}$} is reached, the edge to $l_1$ is taken immediately upon which the value of \out is updated to \intwo.
Similarly, once the expiration time for \randomclock{$\mathit{switch2}$} is reached, the edge back to $l_0$ is taken.
To catch discrete updates of the two input variables, self-loops are added that process this information.
The delay kernel is defined as $\Psi_{e_0} (\sigma)\sim \simulink{Dist}_1$,$\Psi_{e_1}(\sigma)\sim \simulink{Dist}_2$ with $ \simulink{Dist}_i=\mathcal{U}(\simulink{low}, \simulink{high}) \mid \mathcal{N}_{\geq 0} ~(\simulink{var}) $  for $i \in\{1,2\}$ and all states $\sigma \in \Sigma$. The initial state is given by $\Init(l_{0}) = (\{\out= \inone, \randomclock{\mathit{switch1}} := 0, \randomclock{\mathit{switch2}} := 0\}, \texttt{true})$.

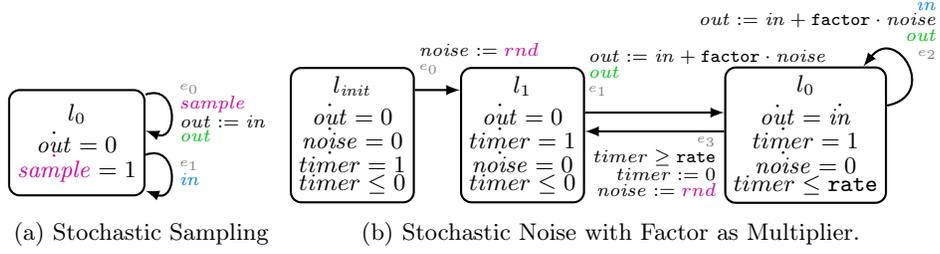
\begin{figure}[tb]
    \centering
    \begin{subfigure}[t]{0.3\textwidth}    
   \centering
    \begin{tikzpicture}
    \node[location, minimum height = 1.4cm] (l0) {$l_0$ \\ [\spacer] $\innervar{\dot{\out}}  =\outervar{0}$ \\ $\randomclock{\dot{\mathit{sample}}} = 1 $ };
    
    \draw[shatransition, sync, bend left = 110, looseness = 2.5] (l0.35)
    to node[right,transition,xshift = 0.cm,yshift=-0cm, align = left] {{\tiny \color{gray} $e_0$} \\
    $\randomclock{\mathit{sample}}$ \\ 
    $\innervar{\out} := \outervar{\innone}$ \\ 
    $\send{\out}$ } 
    (l0.10);
    
    \draw[shatransition, sync, bend left = 110, looseness = 2.5] (l0.-10) 
    to node[right,transition,xshift = 0.cm,yshift=0.cm, align = left]
    {{\tiny \color{gray} $e_1$} \\
    $\receive{\innone}$ }  
    (l0.-35);

\end{tikzpicture}
    \caption{Stochastic Sampling }
    \label{fig:sha-sampling}
    \end{subfigure}%
    ~ 
    \begin{subfigure}[t]{0.7\textwidth}
\centering
    \begin{tikzpicture}
    \node[location] (l0) {$l_0$ \\ [\spacer] $\innervar{\dot{\out}}  =\outervar{\dot{\innone}}$ \\ $\innervar{\dot{timer} = 1}$\\ $\dot{\mathit{noise}} = 0$ \\ $timer \leq \simulink{rate}$};
    
    \node[location, left = 1.85cm of l0] (l1) {$l_1$ \\ [\spacer] $\innervar{\dot{\out}}  = 0$ \\ $\innervar{\dot{timer} = 1}$\\$\dot{\mathit{noise}} = 0$\\ $timer \leq 0$};
    
    \node[location, left =  0.6cm of l1] (init) {$l_{\mathit{init}}$ \\ [\spacer] $\innervar{\dot{\out}}  = 0$\\ $\dot{\mathit{noise}} = 0$ \\ $\innervar{\dot{timer} = 1}$ \\ $timer \leq 0$};
     
    \draw[shatransition, sync, bend right = 110, looseness = 3] (l0.20) 
    to node[right,transition,xshift = -2.9cm,yshift=0.5cm, align = right] {
    $\receive{\innone} $ \\
    $\innervar{\out} := \outervar{\innone} + \simulink{factor}\cdot\innervar{\mathit{noise}} $\\ 
    $\send{\out}$    \\[0.75mm]
    {\tiny \color{gray}$e_2$}} (l0.50);

    \draw[shatransition, sync] ([yshift = 0.05cm]l0.west) 
    to node[below,transition, yshift=0.05cm, xshift=0cm,align = right] {        
    {\tiny \color{gray}$e_3$}\\
    $ \innervar{timer} \geq \simulink{rate}$ \\ 
    $\innervar{timer} := 0$ \\
    $\innervar{\mathit{noise}} := \randomclock{rnd}$ 
    }  ([yshift = 0.05cm]l1.east);

    \draw[shatransition, sync] ([yshift = 0.3cm]l1.east)
    to node[above,transition, yshift=0.1cm, xshift = 0.7cm, align = left] {    
    $ \innervar{\out} := \outervar{\innone} + \simulink{factor} \cdot \innervar{\mathit{noise}} $ \\
    $\send{\out}$    \\[0.75mm]
    {\tiny\color{gray}$e_1$}
    }  
    ([yshift = 0.3cm]l0.west);

       \draw[shatransition, sync] ([yshift= 0.6cm]init.east) to node[above,transition, align = left,yshift=0.1cm, xshift = .55cm] 
       { $\mathit{noise} := \randomclock{rnd}$  \\[0.75mm]
       {\tiny \color{gray}$e_0$}
        } 
     ([yshift= 0.6cm]l1.west);

\end{tikzpicture}
    \caption{Stochastic Noise with Factor as Multiplier.}
    \label{fig:sha-noise}
    \end{subfigure}%
    \caption{SHA Templates for Sampling and Noise.}
\end{figure}

\paragraph{Stochastic Sampling.}
To formalize the stochastic sampling subsystem (see. \allowbreak \Cref{fig:simulink-delay}), we define the SHA template shown in \Cref{fig:sha-sampling}.
The variable \out represents the output of the subsystems, the random clock $\randomclock{\mathit{sample}}$ effectively models the random timer.
The expiration time of $\randomclock{\mathit{sample}}$ is sampled from the distribution specified in the mask.
Once the expiration time is reached the stochastic edge is taken which causes \out to be updated to the current value of \innone.
This discrete update is sent and a new expiration time for $\randomclock{\mathit{sample}}$ is sampled.
As the subsystem has one input, discrete updates of the corresponding variable \innone have to be received, however, this does not affect the value of \out or $\randomclock{\mathit{sample}}$ and does not result in a location change. 
The delay kernel is defined as $\Psi_{e_0}(\sigma)\sim \simulink{Dist} $ with $ \simulink{Dist}=\mathcal{U}(\simulink{low}, \simulink{high}) \mid \mathcal{N}_{\geq 0} ~(\simulink{var}) $ for all states $\sigma \in \Sigma$. 
The initial state is given by $\Init(l_{\mathit{init}}) = (\{\out= \innone,\randomclock{\mathit{sample}} := 0\}, \texttt{true})$.

\paragraph{Stochastic Noise.} 
To formalize the stochastic noise subsystem (see \Cref{fig:simulink-noise}), we define two SHA template: one where the noise is multiplied with a constant factor and one where the rate of distortion depends on the input signal.
Exemplary, we show the SHA template for the former in \Cref{fig:sha-noise}. 
The value of the variable representing the output, i.e. \out, is initially set to the initial value of the input plus a randomly-distributed noise.
Again, we use an initial location to avoid sampled values in the initial state.
\out then evolves with the same rate as the input and every \simulink{rate} time units,  a new value for the variable $\mathit{noise}$ is sampled according to a normal distribution.
As the resets on transitions are non-deterministic, we use an urgent location $l_1$ to ensure that the reset of \out, where the noise is added to the current value of the input signal \innone, is executed after sampling the noise.
Discrete changes of the input variable \innone are handled at the self-loop of $l_0$ by updating the value of \out accordingly.
The reset kernel is given by $\Psi^R_{\mathit{noise}} (\sigma,e_j)\sim \mathcal{N}(\simulink{mean},\simulink{var})  $ for $j \in \{0,3\} $ and all states $\sigma$.
The initial state is given by $\Init(l_{\mathit{init}}) = (\{\out= 0,\mathit{timer} := 0,\mathit{noise} := 0\},  \texttt{true})$. For the SHA template where the rate depends on the input signal see \Cref{app:templates}, \Cref{fig:sha-noiseWithIn}.
\begin{figure}[tb]
\begin{subfigure}[t]{\textwidth}
    \centering
    \begin{tikzpicture}
    \node[location] (aging) {$\mathit{aging}$ \\[\spacer] $\innervar{\dot{\out}}  =\outervar{\dot{\innone}} \cdot (1 - \innervar{age} \cdot \simulink{factor})$\\ $\innervar{\dot{clock}} = 0$\\ $\innervar{\dot{age}} = 0$\\ $\randomclock{\dot{\fail}} = 1 $\\ $\randomclock{\dot{\repair}} = 0 $};    
    \node[location, right=1.2cm of aging] (repairing) {$\mathit{repairing}$ \\[\spacer] $\innervar{\dot{\out}}  =\outervar{\dot{\innone}} \cdot \simulink{rate}_r $\\ $\innervar{\dot{clock}} = 0$\\ $\innervar{\dot{age}} = 0$\\ $\randomclock{\dot{\fail}} = 0 $\\ $\randomclock{\dot{\repair}} = 1 $}; 

    \node[location, left = 2.1cm of aging] (fail){$\mathit{fail}$ \\[\spacer] $\innervar{\dot{\out}}  = 0 $\\ $\innervar{\dot{clock}} = 1$\\ $\innervar{\dot{age}} = 0$\\ $\randomclock{\dot{\fail}} = 0 $\\ $\randomclock{\dot{\repair}} = 0 $\\ $\innervar{clock} \leq 0$};

    \draw[shatransition,sync, bend left = 100, looseness = 1.5] (aging.55) to node[transition, xshift= 0.4cm,yshift=0.25cm, align = left] {
    $\receive{\innone} $\\
    $\innervar{\out}:= \outervar{\innone} \cdot (1 - \innervar{age} \cdot \simulink{factor})$\\
    $\send{\out}$\\[0.75mm]
     {\tiny\color{gray} $e_0$}
    }  (aging.15);

    \draw[shatransition, sync,bend left = 100, looseness = 1.5] (repairing.80) to node[left,transition,xshift= 0.8cm,yshift=0.3cm, align  =left] {
    $\receive{\innone}$\\
    $\innervar{out} := in \cdot \simulink{rate}_r$\\
    $\send{out}$\\[0.75mm]
    {\tiny\color{gray} $e_1$}
    }  (repairing.25);

     \draw[shatransition, sync, bend left = 90, looseness = 2] (fail.100) to node[left,transition,xshift = -0.3cm,yshift=-0.1cm] {
    $\receive{\innone}$ \\[0.5ex]
    {\tiny\color{gray} $e_2$}}  (fail.70);

    \draw[shatransition, sync,] ([yshift=-0cm]repairing.west) to node[below, transition, align = right,xshift=0cm]{
    {\tiny\color{gray} $e_3$}\\[0.5ex]
    $\randomclock{\repair}$\\
    $\innervar{age} := 0$\\
    $\innervar{\out} := \outervar{\innone}$\\
    $\send{\out}$
    } ([yshift=-0cm]aging.east);

    \draw[shatransition,sync,] ([yshift=0.37cm]aging.west) to node[below, transition, align= right, xshift = 0.4cm, align = right] {       
    {\tiny\color{gray} $e_4$}\\
    $\randomclock{\fail}$\\
    $\innervar{clock} := 0$
    }  ([yshift=0.37cm]fail.east);

    \draw[shatransition,sync,] ([yshift=0.55cm]fail.east) to node[above, transition, xshift= 1.1cm,yshift=0.cm, align = left] {
    $\innervar{age} < \simulink{steps} - 1$\\
    $\innervar{\out}:= \outervar{\innone} \cdot (1 - (\innervar{age} + 1) \cdot \simulink{factor})$\\
    $\innervar{age} := \innervar{age} + 1$\\
    $\send{\out}$    \\
    {\tiny\color{gray} $e_5\phantom{e^0}$}
    }  ([yshift=0.55cm]aging.west);

    \draw[shatransition,sync] ([yshift=0cm]fail.south) to node[above right, transition,xshift = 0.8cm, yshift = -0.3cm, align= left] {  
    $age \geq \simulink{steps} - 1$\\
    $\innervar{\out} := \outervar{\innone} \cdot \simulink{rate}_r$\\
    $\send{\out}$\\[0.5ex]
    {\tiny\color{gray} $e_6$}} ([yshift=-0.5cm]fail.south) -| (repairing.south);

\end{tikzpicture}
    \caption{Stochastic Discrete Aging.}
    \label{fig:sha-discreteaging}
\end{subfigure}
~
\begin{subfigure}[t]{\textwidth}
    \centering
    \begin{tikzpicture}
    \node[location, minimum width = 3.8cm] (aging) 
    {$aging$ \\
    [\spacer] $\innervar{\dot{\out}}  =\outervar{\dot{\innone}} \cdot \innervar{\mathit{age}} + \simulink{rate}_a \cdot \outervar{\innone}$ \\ 
    $\innervar{\dot{\mathit{age}}} = \simulink{rate}_a$ \\ 
    $\innervar{\dot{\mathit{clock}}} = 0$\\
    $\randomclock{\dot{\mathit{pause}}} = 1 $ \\ 
    $\randomclock{\dot{\mathit{resume}}} = 0 $ \\ 
    $\randomclock{\dot{\mathit{repaired}}} = 0 $ \\ 
    $\innervar{\mathit{age}} > \simulink{bound}$ 
    };
    
    \node[location, below=5ex  of aging, minimum width = 3.8cm] (repairing) 
    {$repairing$ \\
    [\spacer] $\innervar{\dot{\out}}  = \innervar{\dot{\innone} \cdot \simulink{rate}_r}$ \\
    $\innervar{\dot{\mathit{age}}} = 0$ \\ 
    $\innervar{\dot{\mathit{clock}}} = 0$\\
    $\randomclock{\dot{\mathit{pause}}} = 0 $ \\
    $\randomclock{\dot{\mathit{resume}}} = 0 $ \\
    $\randomclock{\dot{\mathit{repaired}}} = 1 $
    };
    
    \node[location,right=2.3cm of aging, minimum width = 3cm] (pause)
    {$pause$ \\ 
    [\spacer] $\innervar{\dot{\out}}  = 0$ \\ $\innervar{\dot{\mathit{age}}} = 0$ \\ 
    $\innervar{\dot{\mathit{clock}}} = 1$\\
    $\randomclock{\dot{\mathit{pause}}} = 0 $ \\ 
    $\randomclock{\dot{\mathit{resume}}} = 0 $ \\ 
    $\randomclock{\dot{\mathit{repaired}}} = 0 $\\
    $\innervar{\mathit{clock}} \leq 0$
    };
    
    \node[location, minimum width = 3cm] (paused) at (pause|-repairing)
    {$paused$ \\ 
    [\spacer] $\innervar{\dot{\out}}  = \innervar{\dot{\innone} \cdot \innervar{\mathit{age}}}$ \\ $\innervar{\dot{\mathit{age}}} = 0$ \\ 
    $\innervar{\dot{\mathit{clock}}} = 0$\\
    $\randomclock{\dot{\mathit{pause}}} = 0 $ \\ 
    $\randomclock{\dot{\mathit{resume}}} = 1 $ \\ 
    $\randomclock{\dot{\mathit{repaired}}} = 0 $
    };

     \draw[shatransition,sync,] ([yshift=0.5cm]aging.east) to node[above,yshift = -0.1cm,xshift =-0.45cm, transition, align = left] 
     {
     $\randomclock{\mathit{pause}}$ \\
     $\innervar{\mathit{clock}} := 0$ \\[0.5mm]
      {\tiny \color{gray} $e_0$}
     }  ([yshift=0.5cm]pause.west);

    \draw[shatransition,sync,] ([xshift=0.2cm]pause.south) to node[left, transition, align = right] 
    { 
     {\tiny \color{gray} $e_1$}\\$\innervar{\mathit{age}} > \simulink{bound}$
    }  ([xshift=.2cm]paused.north);
    
    \draw[shatransition, sync,] ([yshift=-0cm]paused.west) -- ([xshift=-0.1cm]paused.west) -- 
    ([xshift = 1.9cm,yshift=0cm]aging.east) --  node[below, transition, align = left, xshift = 0cm] {
     {\tiny \color{gray} $e_2$}\\$\randomclock{\mathit{resume}}$ \\
    $\innervar{\out} := \outervar{\innone} \cdot \innervar{\mathit{age}}$ \\
    $\send{\out}$
    } ([yshift=0cm]aging.east);
    
     \draw[shatransition, sync,] ([xshift=-1.75cm]aging.south) to node[left, transition, align = right, near start, yshift=-1mm, xshift = -0.1cm] 
     {
     {\tiny \color{gray} $e_3$}\\
     $\innervar{\mathit{age}} \leq \simulink{bound}$\\
     $\innervar{out} := \outervar{\innone} \cdot \simulink{rate}_r$\\
    $\send{out}$
     }  ([xshift=-01.75cm]repairing.north);
     
     \draw[shatransition, sync,] ([xshift=0cm]pause.west) -- 
      ([xshift=-0.1cm]pause.west)
     --
      ([xshift=1.9cm]repairing.east) --node[ below,transition, align = left,xshift = 0.1cm, yshift = 0cm] 
     {
     {\tiny \color{gray} $e_4$}\\
     $\innervar{\mathit{age}} \leq \simulink{bound}$\\
     $\innervar{out} := \outervar{\innone} \cdot \simulink{rate}_r$ \\
    $\send{\out}$
     }  (repairing.east);
     
    \draw[shatransition, sync,] ([xshift=1.75cm]repairing.north) to node[xshift =0.1cm,yshift = -0.2cm,right, transition, align= left] {
    {\tiny \color{gray} $e_5\phantom{e^0}$}\\
     $\randomclock{\mathit{repaired}}$ \\ 
    $\innervar{\out} := \outervar{\innone}$ \\
    $\innervar{\mathit{age}} := 1$\\
    $\send{out}$\\
    }  ([xshift=1.75cm]aging.south);

    \draw[shatransition, sync, bend right = 110, looseness = 2] (aging.170) to node[near start,transition,xshift = -0.65cm,yshift=0.55cm, align= right] {
     $\receive{\innone} $ \\
     $\innervar{\out} := \outervar{\innone} \cdot \innervar{\mathit{age}} $   \\
     $\send{\out}$\\[0.5mm]
     {\tiny \color{gray} $e_6$}}  (aging.190);
    
    \draw[shatransition,sync, bend left = 110, looseness = 2] (paused.40) to node[near start, transition, xshift=-0.3cm,yshift=0.2cm, align = right]
    {
    $\receive{\innone}$\\
    $\innervar{\out} := \outervar{\innone} \cdot \innervar{\mathit{age}} $\\ 
    $\send{\out}$\\[0.5mm]
    {\tiny \color{gray} $e_7$}
    }  (paused.20);
    
    \draw[shatransition,sync, bend left = 110, looseness = 2.2] (repairing.180) to node[near start,transition,xshift = -0.75cm,yshift=-0.55cm, align= right]
    {
    {\tiny \color{gray} $\phantom{e^0}e_8$}\\
    $\receive{\innone}$\\
    $\innervar{\out} := \outervar{\innone} \cdot \simulink{rate}_r $\\ 
    $\send{\out}$
    }  (repairing.160);

    \draw[shatransition,sync, bend right = 110, looseness = 2] (pause.10) to node[near start, transition, xshift= 0.1cm,yshift=-0.35cm, align = left] {
     {\tiny \color{gray} $e_9\phantom{e^0}$}\\$\receive{\innone} $}  (pause.30);
\end{tikzpicture}
    \caption{Stochastic Continuous Aging.}
    \label{fig:sha-continuousaging}
\end{subfigure}
\caption{SHA Templates for Aging without Explicit Repair Signal.}
\end{figure}
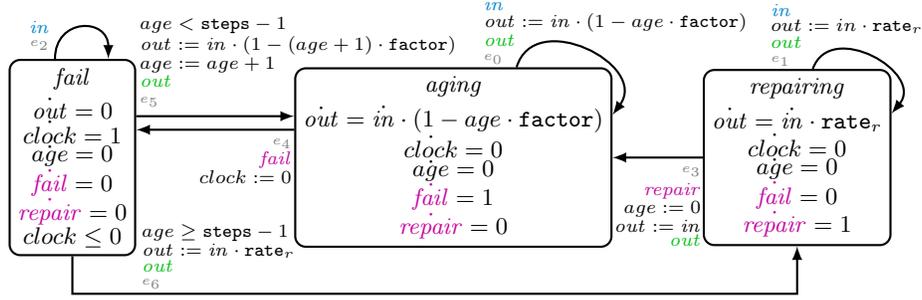
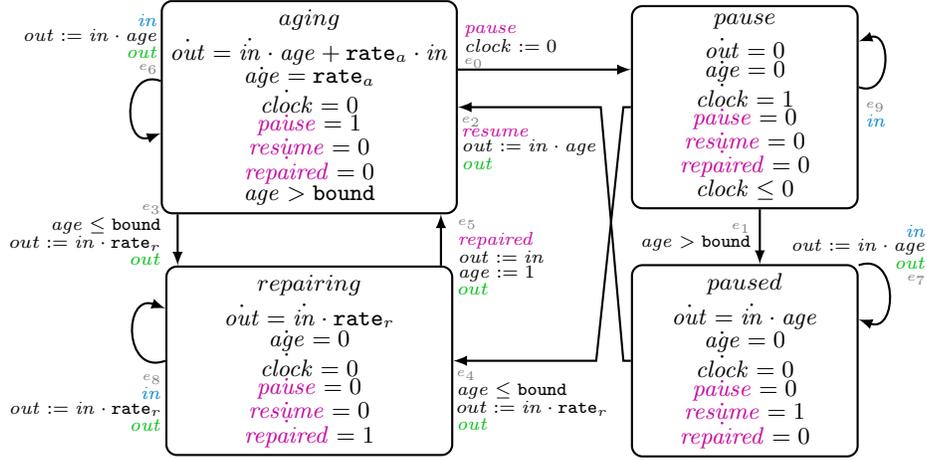
\paragraph{Stochastic Discrete Aging.}
To formalize the stochastic discrete aging subsystem (see \Cref{fig:simulink-agingdiscrete}), we define two SHA templates, one for the case without an explicit repair signal and one with the repair signal. 
Exemplary, we show the SHA template without using an explicit repair signal  in \Cref{fig:sha-discreteaging}. 
The variable \out represents the output of the subsystem which initially equals the value of \innone, i.e. the value of the input signal.
Once the random clock $\randomclock{\mathit{fail}}$ reaches its sampled expiration time, the variable $\mathit{age}$ is increased which controls the rate of the degradation. 
The value of \out is then updated to \innone multiplied by the current aging rate.
Once max degradation is reached, the expiration of $\randomclock{\mathit{fail}}$ triggers the repair of the signal whose duration depends on the expiration time of $\randomclock{\mathit{repair}}$. 
Once the signal is repaired, \out is assigned the value of \innone without degradation.
Discrete changes of the input signal are handled by the self-loops.
The delay kernel is defined as $\Psi_{e_3} (\sigma)\sim \simulink{Dist}_1$,$\Psi_{e_4}(\sigma)\sim \simulink{Dist}_2$ with $ \simulink{Dist}_i=\mathcal{U}(\simulink{low}, \simulink{high}) \mid \mathcal{N}_{\geq 0} ~(\simulink{var}) $  for $i \in\{1,2\}$ and all states $\sigma$.
The initial state is given by $\Init(l_{\mathit{aging}}) = (\{\out= \innone,\randomclock{\mathit{fail}} := 0,\randomclock{\mathit{repair}} := 0, \mathit{age} = 0, \mathit{clock} = 0\}, \texttt{true})$.
The SHA template using an explicit repair signal is shown in \Cref{app:templates}, \Cref{fig:sha-discreteaging2input}. 
\paragraph{Stochastic Continuous Aging.}
To formalize the stochastic continuous aging subsystem (see \Cref{fig:simulink-agingcontinuous}), we again define two SHA templates.
Exemplary, we show the SHA template for the case without using an explicit repair signal shown in \Cref{fig:sha-continuousaging}. 
Similarly to the discrete aging, the variable \out represents the output of the subsystems which is degraded over time until a lower bound is reached.
In contrast to the discrete aging, the rate of the degradation is not increased at discrete time points but reduces continuously, represented by the variable $\mathit{age}$ that evolves with a specified $\simulink{rate}_a$.
Additionally, the aging can be paused and resumed, which is controlled via the two random clocks.
Again, discrete updates of \innone are received via the self-loops in all locations and handled via updating \out.
The delay kernel is defined as $\Psi_{e_0} (\sigma)\sim \simulink{Dist}_1$, $\Psi_{e_2}(\sigma)\sim \simulink{Dist}_2$, $\Psi_{e_5}(\sigma)\sim \simulink{Dist}_3$ with $ \simulink{Dist}_i=\mathcal{U}(\simulink{low}, \simulink{high}) \mid \mathcal{N}_{\geq 0} ~(\simulink{var}) $ for $i \in\{1,2,3\}$ and all states $\sigma$.
The initial state is given by $\Init(l_{\mathit{aging}}) = (\{\out= \innone\cdot \mathit{age},\randomclock{\mathit{pause}} := 0,\randomclock{\mathit{resume}} := 0,\randomclock{\mathit{repair}} := 0, \mathit{age} = 1, \mathit{clock} = 0\}, \texttt{true})$.
The SHA template using an explicit repair signal is shown in \Cref{app:templates}, \Cref{fig:sha-continuousteaging2input}.

\section{Evaluation}
\label{sec:eval}
To demonstrate the feasibility of our approach,
we provide quantitative results for two small case studies. The first is a modified version of the temperature control system shown in \Cref{fig:examplesimulink},  where sensor losses are modeled as delayed sampling. The second is a simplified energy measurement unit, which stochastically switches between low and high loads. 
To formally analyze both systems, we have used the transformation presented in \cite{blohm2025towardsQuantitative} extended as outlined in \Cref{sec:formalization}.
\begin{figure}[b]
        \centering
    \begin{tikzpicture}

    \node[block,  minimum width = 1.2cm,minimum height = 1.2cm ] (switch) {};
    \draw[-, semithick] ($(switch.west)+(-0cm,0.4cm)$) to  ($(switch) + (-0.1cm,0.4cm)$);
    \draw[draw=black, semithick] ($(switch) + (-0.1cm,0.36cm)$) rectangle ++(0.08,0.08);
    \draw[-, semithick] ($(switch.west) + (0,-0.4cm)$) to ($(switch) + (-0.1cm,-0.4cm)$);
    \draw[draw=black, semithick] ($(switch) + (-0.1cm,-0.44cm)$) rectangle ++(0.08,0.08);    
    \draw[-, semithick] (switch) to node[right, font=\boldmath\scriptsize\bfseries]{$>$\,0} ($(switch.west) + (0.2cm,0)$);
    \draw[-, semithick] ($(switch) + (-0.02cm,0.4cm)$)  to ($(switch.east) + (-0.3cm,0)$)to (switch.east);

    \node[subsystem, minimum width = 1cm, minimum height = 0.8cm,right = 0.3cm of switch] (stochastic) {};
    \node[blocklabel, below =0.1ex of stochastic] (timer2L) {stochastic\\ sampler};
    
   \node[block, right = 0.3cm of stochastic, font=\boldmath] (int) {$\frac{1}{s}$};

   \node[block, left = 1.3cm of switch] (relay) {};
    \draw[-, color=lightgray!70, semithick] ($(relay.north) + (0,-0.04cm)$) to ($(relay.south) + (0,0.04cm)$);
    \draw[-, color=lightgray!70, semithick] ($(relay.west) + (0.04cm,0)$) to ($(relay.east) + (-0.04cm,0)$);
    \draw[-, semithick] ($(relay.west) + (0.04cm, -0.25cm)$) -| ($(relay) + (0.1cm, 0.25cm)$);
    \draw[-, semithick] ($(relay) + (-0.1cm, -0.25cm)$) |- ($(relay.east) + (-0.04cm, 0.25cm)$);    

    \node[block, font=\scriptsize\boldmath] (sum) at ([xshift = -0.6cm, yshift = 0cm]relay.west){$+-$};

    \node[block] (tdes) at ([xshift =-0.6cm, yshift = -0cm] sum.west) {\scriptsize$21$\\ \tiny tdes};
    \node[block] (heat)at  ($(switch.west)+(-0.7cm,0.4cm)$) {\scriptsize 0.03\\ \tiny heat}; 
    \node[block] (cool) at  ($(switch.west)+(-0.7cm,-0.4cm)$) {\scriptsize -0.03\\ \tiny cool};

    \draw[simulinksignal] (tdes) -- (sum);
     \draw[simulinksignal] (switch.east) --(stochastic.west);
     \draw[simulinksignal] (stochastic.east) --(int.west);
     \draw[simulinksignal] (heat.east) -- ([xshift=-0.3cm]switch.west |- heat.east) -- ([xshift=-0.3cm]$(switch.west)+(-0cm,0.4cm)$) --($(switch.west)+(-0cm,0.4cm)$);
     \draw[simulinksignal] (cool.east) -- ([xshift=-0.3cm]switch.west |- cool.east) -- ([xshift=-0.3cm]$(switch.west)+(-0cm,-0.4cm)$) --($(switch.west)+(-0cm,-0.4cm)$);
     \draw[simulinksignal] (sum.east) -- (relay.west);
    \ \draw[simulinksignal] (relay.east) -- (switch.west);
    \draw[simulinksignal] (int.east) -- ([xshift = 0.15cm] int.east) --([xshift = 0.15cm,yshift=-0.5cm] int.south east)  -| ([xshift = 0cm]sum.south);
\end{tikzpicture}
    \caption{Simulink Model for the Temperature Control System with Loss.}
    \label{fig:example-simulink-delayed}
\end{figure}
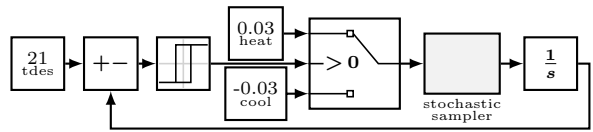
Then, we have used the tool \texttt{modes} \cite{Modes} from the \modest to apply statistical model checking, which provides us with statistical guarantees in the form of confidence intervals (CI). 
We compare the CI provided by \texttt{modes} with a simulation-based evaluation of the Simulink model. 
While Simulink does not provide CI as-is, we use the implementation presented in \cite{isola23} to compute them based on the Wilson score \cite{wilson}.
For both tools we use a confidence level of $\lambda = 0.95$ and the Wilson score to compute the CI.
We use a time horizon of $t = 100$ and perform 5000 runs in the Simulink model and 9704 for \texttt{modes}, as the significantly faster runtime of \texttt{modes} allow us to perform more runs.
In Simulink a fixed-step solver with a step size $s = 0.05$ is used and for \texttt{modes} either a uniform scheduler (temperature control) or a ASAP scheduler (energy consumption) is used.

\paragraph{Temperature Control System.}
The temperature control system with stochastic sensor loss is shown in \Cref{fig:example-simulink-delayed}. 
It is similar to the example we have used in the introduction to Simulink (cf. \Cref{fig:examplesimulink}). 
To model a sensor with stochastic loss, i.e. 
that the sensor can only successfully read the temperature at stochastically chosen time points, we use a stochastic sampling block.
The 
\begin{wrapfigure}[18]{r}{0.3\textwidth}
\vspace{-0.5\intextsep}
    \begin{tikzpicture}
    \node[location] (l0) {$\mathit{cool}$ \\ [\spacer] $\innervar{\dot{\mathit{temp}}}  =\mathit{rate}$ \\ $\randomclock{\dot{\mathit{sample}}} = 1 $\\ [\spacer] $\mathit{temp} \geq 20.5$};
    \node[location, below=01cm  of l0] (l1) {$\mathit{heat}$ \\ [\spacer] $\innervar{\dot{\mathit{temp}}}  =\mathit{rate}$ \\ $\randomclock{\dot{\mathit{sample}}} =1 $ \\ [\spacer] $\mathit{temp} \leq 21.5$};

    \draw[shatransition,sync,] ([xshift=0.1cm]l0.south) to node[right, transition, yshift = 0.0cm,align=left] {   
    {\tiny\color{gray} $e_0\phantom{e^0}$}\\
    $\mathit{temp} \leq 20.5$\\ 
    $ \mathit{val} := 0.03$
    }  
    ([xshift=0.1cm]l1.north);
    
    \draw[shatransition, sync,] ([xshift=-0.1cm]l1.north) to node[left, transition,yshift=-0cm,align=right] 
    {$\mathit{temp} \geq 21.5$\\ 
    $ \mathit{val} :=- 0.03$\\
    {\tiny\color{gray} $\phantom{e^0}e_1$}} 
    ([xshift=-0.1cm]l0.south);
    
    \draw[shatransition, sync, bend right = 100, looseness = 1.7] (l0.20) to node[right,transition,xshift = -0.8cm,yshift=0.3cm, align = right] {$\randomclock{\mathit{sample}}$ \\$\mathit{rate} :=\mathit{val} $\\{\tiny\color{gray} $\phantom{e^0}e_2$} }  (l0.60);
    
    \draw[shatransition,sync, bend right = 100, looseness = 1.7] (l1.-160) to node[left, transition, xshift= 0.65cm,yshift=-0.15cm, align =left] {{\tiny\color{gray} $e_3\phantom{e^0}$}\\$\randomclock{\mathit{sample}}$ \\$\mathit{rate} :=\mathit{val} $}  (l1.-120);    
\end{tikzpicture}
    \vspace{-0.8\intextsep}
\caption{Temperature Control as \hawk.}
    \label{fig:example-sha-tempControl}
\end{wrapfigure}
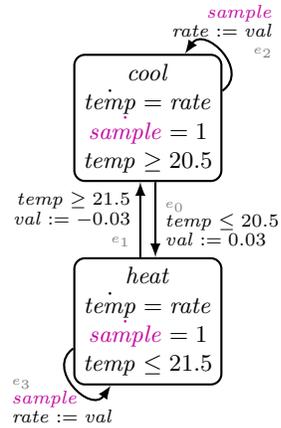
corresponding \hawk is shown in \Cref{fig:example-sha-tempControl}. Please note that we have applied some optimizations to eliminate redundant or unused variables, locations and edges. 
The delay kernel 
is defined as $\Psi_{e_2} = \Psi_{e_3} (\sigma)\sim \mathcal{U}(10,20)$ for all states $\sigma$ and $\Init(l_{\mathit{cool}}) = \{\mathit{temp}= 21,\randomclock{\mathit{sample}} := 0,\mathit{rate} = -0.03\}$.
\begin{table}[tb]
    \caption{Results for Temperature Control System with Sensor Loss.}
    \label{tab:results-delay}
    \centering
    \begin{tabular}{c|c|c|c|c|c}
         \multicolumn{2}{c|}{\textbf{Tool} }& $P( \Diamond~ \mathit{tmp} \leq 20)$ &  $P( \Diamond~\mathit{tmp} \leq 20.2)$ & $P( \Diamond~\mathit{tmp} \leq 20.4)$  &   $P( \Diamond~\mathit{tmp} \leq 20.5)$\\ \hline \hline
         \multirow{3}{*}{\rotatebox[origin=t]{90}{\parbox{0.55cm}{\centering\scriptsize Simulink}}} &CI & $[0.0652, 0.0795]$ &$ [0.4454, 0.4730]$ & $[0.7262, 0.7506]$&  $[0.9992, 1]$ \\
         &midpoint &$0.0723$& $0.4592$& $0.7384$& $0.9996$ \\
         \hline
       \multirow{3}{*}{\rotatebox[origin=t]{90}{\parbox{0.2cm}{\centering \scriptsize\texttt{modes}}}} & CI & $[0.0497, 0.05885]$ & $[0.4463, 0.4663]$ &  $[0.7322, 0.7498]$ & $[0.9997, 1]$  \\
         &midpoint&$0.0541$&$0.4562$&$0.7409$& $1$\\
    \end{tabular} 
\end{table}
We use a PCTL-like notation to express the properties that a temperature stays below a given threshold, e.g., $P( \Diamond~ \mathit{temp} \leq 20)$ gives the probability that a $\mathit{temp}$ of $20$ or lower is reached during the observed time frame.
\Cref{tab:results-delay} shows that the CI provided by \texttt{modes} are tighter and lie within the CI computed with the Simulink model.
As expected, a temperature of $20.5$ is reached almost certainly, whereas, a temperature of $20$ is quite unlikely, as the controller aims to keep the temperature at $21$ degrees.
On average, the analysis performed with \texttt{modes}  was significantly faster with only $0.3s$, whereas the Simulink evaluation exceeded 30 minutes ($1802s$). 

\begin{figure}[b]
    \centering
    \begin{tikzpicture}
    \node[subsystem, minimum width = 1.1cm, minimum height =1cm] (switch1) {};
    \node[blocklabel, below =0.1ex of switch1] (switch1L) {st. switch 1};
    \node[] (switch1In1) at ([yshift=0.4cm]switch1.west){};
    \node[] (switch1In2) at ([yshift=-0.4cm]switch1.west){};
    
    \node[subsystem, minimum width = 1.1cm, minimum height =1cm, right = 1.4cm of switch1] (switch2) {};
    \node[blocklabel, below =0.1ex of switch2] (switch2L) {st. switch 2};
    \node[] (switch2In1) at ([yshift=0.4cm]switch2.west){};
    \node[] (switch2In2) at ([yshift=-0.4cm]switch2.west){};

    \node[block] (zero1)  at ([xshift=-0.6cm]switch1In1) {\scriptsize$0$};
    \node[block] (zero2) at ([xshift=-0.6cm]switch2In1) {\scriptsize$0$};
    \node[block] (three) at ([xshift=-0.6cm]switch1In2)  {\scriptsize$300$};
    \node[block] (one) at ([xshift=-0.6cm]switch2In2) {\scriptsize$100$};
    
    \node[block, font=\scriptsize] (sum) at ([xshift=0.8cm, yshift=-0cm] switch2.east) {$++$};
    
    \node[block,font=\scriptsize, minimum height = 0.4] (comp) at ([xshift =4cm, yshift = -0cm]zero2){$\geq 400$};
    \node[block,font=\boldmath, right= 0.4cm of comp] (int1) {$\frac{1}{s}$};
    \node[blocklabel, below =0.1ex of int1] (int1L) {max load};
    \node[block,font=\boldmath, below=0.1cm of comp] (int2) {$\frac{1}{s}$};
    \node[blocklabel, below =0.1ex of int2] (int2L) {total};
    
    \draw[simulinksignal] (zero1) -- (switch1In1.center);
    \draw[simulinksignal] (zero2) -- (switch2In1.center);
    \draw[simulinksignal] (three) -- (switch1In2.center);
    \draw[simulinksignal] (one) -- (switch2In2.center);
    \draw[simulinksignal] (switch1.east) -| ([xshift=0.2cm,yshift=-0.8cm]switch1.east) -| ([xshift = -0.3cm,yshift=-0.2cm]sum.west) --([yshift=-0.2cm]sum.west);
    \draw[simulinksignal] ([yshift=0.2cm]switch2.east|-sum.west) -- ([yshift=0.2cm]sum.west);
    \draw[simulinksignal] (sum.east)-- ([xshift=0.1cm]sum.east) node[dot] {} |- (comp.west);
    \draw[simulinksignal] ([xshift=0.1cm]sum.east) |- (int2.west);
    \draw[simulinksignal] (comp) -- (int1);
    
\end{tikzpicture}
    \caption{Simulink Model for the Energy Measurement System.}
    \label{fig:example-simulink-switch-energy}
\end{figure}
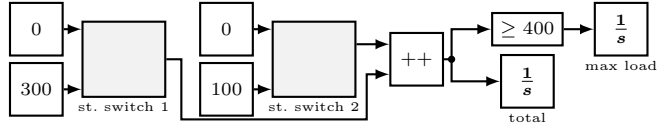
\begin{table}[tb]
    \caption{Results for the Energy Measurement Unit.}
    \label{tab:results-switching}
    \centering
    \begin{tabular}{c|c|c|c|c|c}
         \multicolumn{2}{c|}{\textbf{Tool} }& $P( \Diamond~ \mathit{load} \geq 10)$ &  $P( \Diamond~\mathit{load} \geq 30)$ & $P( \Diamond~\mathit{total} \geq 13000)$  &   $P( \Diamond~\mathit{total} \geq 16000)$\\ \hline \hline
         \multirow{3}{*}{\rotatebox[origin=t]{90}{\parbox{0.55cm}{\centering\scriptsize Simulink}}} &CI & $[0.8934, 0.9099]$ &$ [0.0000, 0.0008]$ & $[0.8895, 0.9063]$&  $[0.0050, 0.0097]$ \\
         &midpoint &$0.9017$& $0.0004$& $0.8979$& $0.0074$ \\
         \hline
       \multirow{3}{*}{\rotatebox[origin=t]{90}{\parbox{0.2cm}{\centering \scriptsize\texttt{modes}}}} & CI & $[0.8975, 0.9094]$ & $[0, 0.0005]$ &  $[0.8914, 0.9036]$ & $[0.0043, 0.0074]$  \\
         &midpoint&$0.9034$&$0$&$0.8975$& $0.0057$
    \end{tabular} 
\end{table}

\begin{wrapfigure}[16]{r}{0.47\textwidth}
\vspace{-0.85\intextsep}
    \begin{tikzpicture}
    \node[location] (l0) {$l_0$ \\ 
    $\dot{\mathit{total}} =0$ \\ $\dot{\mathit{max}} =0$\\
    $\randomclock{\dot{\mathit{switch_{1,1}}}} = 1 $\\ 
    $\randomclock{\dot{\mathit{switch_{1,2}}}} = 0 $\\$\randomclock{\dot{\mathit{switch_{2,1}}}} = 1$\\ $\randomclock{\dot{\mathit{switch_{2,2}}}} = 0$};
    
    \node[location, right=1.3cm  of l0] (l1) {$l_1$ \\  $\dot{\mathit{total}} =100$ \\ $\dot{\mathit{max}} =0$ \\$\randomclock{\dot{\mathit{switch_{1,1}}}} = 0 $\\ 
    $\randomclock{\dot{\mathit{switch_{1,2}}}} = 1 $\\ $\randomclock{\dot{\mathit{switch_{2,1}}}} = 1 $ \\$\randomclock{\dot{\mathit{switch_{2,2}}}} = 0 $};
    
    \node[location, below = 0.8cm of l0] (l2) {$l_2$ \\ $\dot{\mathit{total}} =300$ \\ $\dot{\mathit{max}} =0$ \\$\randomclock{\dot{\mathit{switch_{1,1}}}} = 1 $\\ 
    $\randomclock{\dot{\mathit{switch_{1,2}}}} = 0 $\\ $\randomclock{\dot{\mathit{switch_{2,1}}}} = 0 $\\$\randomclock{\dot{\mathit{switch_{2,2}}}} = 1 $};
    
    \node[location, right=1.3cm  of l2] (l3) {$l_3$ \\  $\dot{\mathit{total}} =400$ \\ $\dot{\mathit{max}} =1$ \\$\randomclock{\dot{\mathit{switch_{1,1}}}} = 0 $\\ 
    $\randomclock{\dot{\mathit{switch_{1,2}}}} = 1 $\\ $\randomclock{\dot{\mathit{switch_{2,1}}}} = 0 $\\$\randomclock{\dot{\mathit{switch_{2,2}}}} = 1$};

    \draw[simulinksignal] ([yshift=0.3cm]l0.east) -- node[transition, above] {$\randomclock{\mathit{switch_{1,1}}}$\\{\tiny\color{gray} $e_0$}} ([yshift=0.3cm]l1.west);
    \draw[simulinksignal] ([yshift=-0.3cm]l1.west) -- node[transition, below] {{\tiny\color{gray} $e_1$}\\$\randomclock{\mathit{switch_{1,2}}}$} ([yshift=-0.3cm]l0.east);
    
    \draw[simulinksignal] ([yshift=0.3cm]l2.east) -- node[transition, above] {$\randomclock{\mathit{switch_{1,1}}}$\\{\tiny\color{gray} $e_2$}} ([yshift=0.3cm]l3.west);
    \draw[simulinksignal] ([yshift=-0.3cm]l3.west) -- node[transition, below] {{\tiny\color{gray} $e_3$}\\$\randomclock{\mathit{switch_{1,2}}}$} ([yshift=-0.3cm]l2.east);

    \draw[simulinksignal] ([xshift=0.1cm]l0.south) -- node[transition, left, align=right] {$\randomclock{\mathit{switch_{2,1}}}$\\{\tiny\color{gray} $e_4$}} ([xshift=0.1cm]l2.north);
    \draw[simulinksignal] ([xshift=0.3cm]l2.north) -- node[transition, right,align=left] {{\tiny\color{gray} $e_5$}\\$\randomclock{\mathit{switch_{2,2}}}$} ([xshift=0.3cm]l0.south);

    \draw[simulinksignal] ([xshift=-0.3cm]l1.south) -- node[transition, left, align=right] {$\randomclock{\mathit{switch_{2,1}}}$\\{\tiny\color{gray} $e_6$}} ([xshift=-0.3cm]l3.north);
    \draw[simulinksignal] ([xshift=-0.1cm]l3.north) -- node[ transition, right, align=left] {{\tiny\color{gray} $e_7$}\\$\randomclock{\mathit{switch_{2,2}}}$} ([xshift=-0.1cm]l1.south);
\end{tikzpicture}
    \caption{Energy Measurement Unit as a \hawk.}
    \label{fig:example-sha-switch-energy}
\end{wrapfigure}
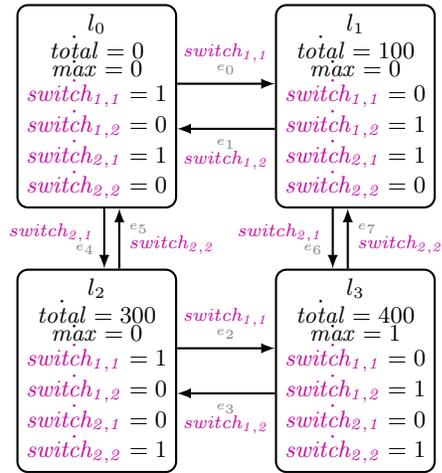

\paragraph{Energy Measurement Unit.}
The second case study is inspired by an energy measurement  unit.
The Simulink model shown in \Cref{fig:example-simulink-switch-energy} uses two stochastic switches that both model a unit that can stochastically switch between a high or low load.
Two integrators are used to measure the total energy consumed and the amount of time passed with a maximum load, i.e. with both consumers in high load mode.
The corresponding \hawk is shown in \Cref{fig:example-sha-switch-energy}.
The four different locations reflect the four different states of energy consumption:
 1) both units have a low load($l_0$) 2) only one unit has high load ($l_1$ and $l_2$) or 3) both units have high load ($l_3$).
The delay kernel is defined as $\Psi_{e_0} = \Psi_{e_2} (\sigma)\sim \mathcal{U}(10,20)$, $\Psi_{e_1} = \Psi_{e_3} = \Psi_{e_4} = \Psi_{e_6} (\sigma)\sim \mathcal{U}(5,10)$, and $\Psi_{e_5} = \Psi_{e_7} (\sigma)\sim \mathcal{U}(5,15)$ for all states $\sigma$ and $\Init(l_0 = \{\mathit{total}= 0,\mathit{max}= 0,\randomclock{\mathit{switch}_{1,1}} = 0,\randomclock{\mathit{switch}_{1,2}} = 0,\randomclock{\mathit{switch}_{2,1}} = 0,\randomclock{\mathit{switch}_{2,2}} = 0\}$.
We analyzed whether the total energy consumption and the time spent at max load exceed certain thresholds.
\Cref{tab:results-switching} shows
that the CI provided by \texttt{modes} are tighter and lie within the CI provided by Simulink for the first three properties.
For the fourth property $P(\Diamond \mathit{total} \geq 16000)$, the CI provided by \texttt{modes} is also tighter but there is only a large overlap.
As the probabilities for this property are very low, we suspect that slight differences in the sampling process might be the cause.
Further investigation is needed to better understand the causes. 
The average runtime using \texttt{modes} was significantly faster with only $0.3s$ while Simulink exceeded 12 minutes ($736s$).

\section{Related Work} 
\label{sec:relwork}
Different works have investigated hybrid systems under uncertainty modeled in Simulink, e.g. \cite{kuriakose2019customized,chiacchio2018dynamic,zuliani2012rare,chen2013simulink}.
In \cite{kuriakose2019customized}, a stochastic timer is used to model uncertain operation times.
\cite{chiacchio2018dynamic} presents a power plant with a failure-repair model, where 
they also consider degradation over time. 
\cite{zuliani2012rare} 
presents a model of a controller for an air-craft elevator system and introduce random failures to three hydraulic circuits based on Poisson processes.
\cite{chen2013simulink} presents a Simulink model of a human heart with stochastic delay between subsequent  heartbeats.
While these approaches model and analyze relevant uncertainties, they 
are rather specific to the used case studies. 
Additionally, all of these works perform simulation within Simulink and are not amenable for a sophisticated quantitative analysis.

There have been quite some efforts to enable the formal verification of hybrid systems  modeled in Simulink, e.g. \cite{chutinan2003computational,minopoli2016sl2sx,zou2015formal,chen2017mars,liebrenz2018deductive,liebrenz2019service}. 
In \cite{chutinan2003computational}, the authors propose the tool CheckMate to model HA in Simulink, which can then be formally verified via reachability analysis.
Similarly, in \cite{minopoli2016sl2sx} the authors present a transformation from a subset of Simulink to the HA dialect SpaceEx~\cite{FrehseKLG13}. 
However, they focus on techniques for a special class of systems and do not provide general transformation rules for a broader set of blocks. 
In \cite{zou2015formal,chen2017mars}, the authors transform Simulink models with Stateflow parts into Hybrid CSP and enable the verification in the Hybrid Hoare Logic Prover. 
Finally, in our own previous work \cite{liebrenz2018deductive,liebrenz2019service,Adelt2021}, we have presented a transformation from Simulink into the differential dynamic logic \cite{platzer2008differential}, which enables deductive verification using KeYmaera~X \cite{fulton2015keymaera}.
Additionally, in \cite{bak2019hybrid} the authors propose a modular correct-by-construction  approach for embedded systems where HA are first modeled and verified and then translated and embedded into Simulink/Stateflow models. 
However, all of these methods focus on the qualitative analysis of safety properties, and none of them take stochastic components 
into consideration. 

To evaluate models with stochastic behavior, SMC approaches for Simulink have been proposed \cite{zuliani2013bayesian,legay2016statistical}. 
While \cite{zuliani2013bayesian} is based on Bayesian statistics and hypothesis testing, \cite{legay2016statistical} uses Plasma lab with Monte Carlo simulation for probability estimation. However, both model uncertainties in an ad hoc manner for a specific scenario 
and rely on expensive simulations in Simulink. Furthermore,
their approach is not amenable to quantitative analysis beyond SMC. 
In \cite{filipovikj2016simulink}, the authors propose a transformation from Simulink into stochastic timed automata that can be analyzed with \uppaal SMC~\cite{david2015uppaal}. 
However, they do not consider stochastic blocks and transform the Simulink models into a deterministic STA model where all probabilities are one.

Our transformation from Simulink into SHA requires a well-defined formalism that can express the hybrid and stochastic behavior present in Simulink.
HA~\cite{alur19953,henzinger1998decidable} have been extended with stochastic components as e.g. stochastic timed automata~\cite{bertrand_stochastic_2014}, or singular and rectangular automata with random events \cite{daSilva23,delicarisTASE23,Delicaris25journal}.
However, these approaches do not allow us to model the complex continuous behavior present in Simulink.
More general classes for stochastic hybrid models are considered in~\cite{lygeros_stochastic_2010}, which even allow for stochastic differential equations to express stochastic noise.
However, to the best of our knowledge, this class is not directly amenable for the computation of reachability probabilities.
\section{Conclusion}
\label{sec:conclusion}
In this paper, we have presented an approach to model uncertainties in Simulink by providing a library of stochastic subsystems.
This library allows us to capture 
uncertainties like noise or aging and model real-life systems more accurately.
We have also presented transformation rules for these subsystems to formally capture their behavior via SHA templates and seamlessly integrated these templates into our previously proposed transformation from Simulink to SHA.
This allows us to formally argue about safety and performance under uncertainty.
To demonstrate the feasibility of our approach, we have presented two small case studies using our newly presented stochastic subsystems. 
We have analyzed different properties using SMC on the transformed SHA with the tool \texttt{modes} as well as an evaluation in Simulink.
Our results show that the analysis using SHA provides tighter CI with only a fraction of the computational effort required for the Simulink evaluation.
Additionally, the SHA are not limited to SMC and could be analyzed with other quantitative analysis techniques 
in the future.

In future work, we plan to apply more sophisticated quantitative analysis techniques like reachability analysis, e.g. using \realyst \cite{delicarisTASE23,Delicaris2024}. Furthermore, we plan to investigate how the existing tools can be extended to analyze more complex, non-linear systems, for example by combining them with 
deductive verification, which can not cope well with stochasticity but might be useful to analyze non-linear differential equations.  
\paragraph{Data Availability.} The models, scripts, and tools to reproduce our experimental evaluation, the Simulink files for the stochastic subsystems and an extended version of the paper including an appendix are archived and publicly available at DOI~\href{https://doi.org/10.5281/zenodo.15273669}{10.5281/zenodo.15273669}. 

\appendix
\label{sec:appendix}
\section{Appendix}
\subsection{Probability Measures and Stochastic Kernels}
\label{app:ProbabilityIntro}

A random \emph{experiment} takes an outcome from a \emph{sample space} $\Omega$, whose subsets are called \emph{events}.
A \emph{$\sigma$-algebra} $\F$ is a set of events which contains the event that corresponds to the complete set $\Omega$ and is closed under complement and countable union. The standard Borel $\sigma$-algebra  $\mathcal{B}(\Omega)$  is the smallest $\sigma$-algebra containing all open events.
An event is \emph{$\F$-measurable} if it is in $\F$.
The pair $(\Omega, \F)$ is called a \emph{measurable space}.

Given $(\Omega, \F)$, a \emph{probability measure} is a function $\Pr : \F \rightarrow [0, 1] \subseteq \R$ with
(i) $\Pr (\Omega) = 1$,
(ii) $\Pr(E)=1-\Pr(\bar{E})$ for all $E \in \F$ and 
(iii) $\Pr (\bigcup_{i=0}^{\infty} E_i ) = \Sigma_{i=0}^{\infty} \Pr (E_i )$ for any $E_i \in  \F$ with $E_i \cap E_j = \emptyset$ for all $i, j \in \N$, $i \not= j$.
%

%

Let $(\Omega, \F)$ and $(S, \Sigma)$ be measurable spaces,
$X : \Omega {\rightarrow} S$, $s {\in} S$ and $\sigma{\in}\Sigma$. We define
$X \sim s$ to be $\{\omega \in\Omega\suchthat X (\omega) \sim s\}$ and
    $X^{-1}(\sigma) = \bigcup_{s\in\sigma} (X \sim s)$ with $\sim\in\{\leq,<,=,>\geq\}$.
$X$ is \emph{measurable} (wrt. $(\Omega, \F)$ and $(S, \Sigma)$) if $X^{-1}(\sigma) \in \F$ for all $\sigma \in\Sigma$.
A \emph{random variable} is a measurable function $X : \Omega \rightarrow S$;
we call $X(\omega)$ the \emph{realization} of $X$ for $\omega\in\Omega$.

For the following we instantiate $\Omega=S=\Rpz$, $\F=2^{\Rpz}$ and $X$ the identity.  For $f : \Rpz \to \Rpz$ we define its \emph{support} as $\support(f) = \{\omega \in \R \suchthat f (\omega) > 0\}$. We require $f$ to be absolute continuous with $\int_{0}^{\infty}  f(\omega) \textit{d}\omega = 1$ and call  $f$  a \emph{probability density function (PDF)}, which induces for all $a\in \Rpz$ the unique probability measure $\Pr:2^{\Rpz}\rightarrow [0,1]$ with $\Pr(X\leq a)=\int_{0}^{a} f(\omega) \textit{d}\omega$, and the \emph{cumulative distribution function} (CDF) $F:\Rpz\to[0,1]$ with $F(a)=\Pr(X\leq a)$.

We denote the set of all continuous probability distributions by $\Distr$. Given two measurable spaces $ (\Omega_1,\F_1) $ and  $ (\Omega_2,\F_2) $, a \emph{stochastic kernel from $ (\Omega_1,\F_1) $ to $ (\Omega_2,\F_2) $} \cite{Klenke2014ProbabilityCourse} is a function $\kappa:\F_2 \times \Omega_1\to[0,1]$ with:
	\begin{itemize}
	\item For each $ E_2\in\F_2 $, the function $f_{E_2}^{\kappa}: \Omega_1\rightarrow [0,1]$ with $f_{E_2}^{\kappa}(\omega_1)=\kappa(E_2,\omega_1) $ is measurable w.r.t. $(\Omega_1, \F_1)$ and $([0,1],\mathcal{B}([0,1])$.
	\item For  each $ \omega_1\in\Omega_1 $, the function $ \Pr^{\kappa}_{\omega_1}: \F_2 \rightarrow [0,1]$ with $\Pr^{\kappa}_{\omega_1}(E_2)=\kappa(E_2,\omega_1) $ is a probability measure on $(\Omega_2,\F_2)$.
	\end{itemize}
        Stochastic kernels are used to express the state-dependent probability $\kappa(E_2,\omega_1)$ of event $E_2\in\F_2$ in system state $\omega_1\in\Omega_1$. 

\subsection{Formal Construction of DHA}
\label{app:constrHEE}
To define the semantics of DHA \cite{willemsen23Comparing,willemsen24camels}, the states $\sigma\in\Sigma$ are extended with the storage of the sampled delays for each label, i.e. the extended states have the form $(\sigma,\realisierungen)\in\States\times\Rpz^k$ for $k=|\Lab|$.

Formally,  $\H=(\Loc, \Var, \allowbreak \Flow, \Inv, \Lab, \Edge, \Init)$ is extended to $\HEEdecomp=(\Loc, \allowbreak \VarEdecomp, \FlowEdecomp, \allowbreak \Inv, \Lab, \EdgeEdecomp, \InitEdecomp)$ with $\VarEdecomp=\Var\cup\{\auxwatch_1,\ldots\auxwatch_k\}$; 
with $\Jumps(\state)=\{a\in \Lab \suchthat \exists \state'\in\States.\ \state\xrightarrow{a}\state'\}$ indicating the set of enabled edges in state $\sigma$, 
$\FlowEdecomp $ extends $ \Flow $  with 
\begin{align*}
\dot{\stopwatchDecomposed{i}}&=\begin{cases}
		1, &\textit{ if } \{e=(\ell,a_i,g,r,\ell')\,|\,e\in\Jumps(\sigma)\} \\
		0, &\textit{ else,}
	\end{cases}
\end{align*}
to track the enabling times.
$\InvEdecomp(\ell)=\Inv(\ell)\times\Rpz^k$; $\InitEdecomp(\ell)=\Init(\ell)\times\{0\}$ with $0\in\Rpz^k$ for all $\ell\in\Loc$; and $\EdgeEdecomp=\{(\ell,a_i,\gEdecomp,\rEdecomp,\ell')\,|\,\allowbreak (\ell,\allowbreak a_i,\allowbreak g,\allowbreak r,\ell')\in\Edge\}$ with $\gEdecomp=g\times\Rpz^k$ and $\rEdecomp(\ass)=(r(\ass),\ass')$ with $\ass'\in\Rpz^k$, $\ass'(\auxwatch_{i})=0$ and $\ass'(\auxwatch_j)=\ass(\auxwatch_j)$ for all $j\in\{1,\ldots,k\}\setminus\{i\}$ (i.e., each edge resets the semantic watch for its label to $0$).

\subsection{Lifting SHA Template Composition to \hawk}
\label{app:transformation}
To define the SHA templates, we introduce \hawksync, which are defined analogously as \LACsync in \cite{blohm2025towardsQuantitative}.
That is, they introduce an additional labeling function to assign \emph{sending} and \emph{receiving} variables to edges and separate the variable set into \emph{input} and \emph{output} variables. Also, they relax the \Flow and the kernels to only consider output variables. Furthermore, \Init is now a pair of state and condition.
\begin{definition}
\label{def:hawksync}
A \hawksync is a tuple $\hawksyncinstance= (\mathcal{H},\kernel,\resetKernel, \Sync)$ with  $\mathcal{H}=(\Loc, \Var, \Flow, \allowbreak\Inv, \Lab, \Edge, \Init)$ where
\begin{itemize}
    \item \Loc, \Inv, $\Lab$, and $\Edge $ are defined as for \hawk.
    \item $\Var= \VarInner \cup \VarOuter $ is the set of variables.
    \item \Init  assigns a pair of initial state and condition to each location $l \in \Loc$ for each $v \in \VarInner$. A condition is a linear formula.
    \item \Flow assigns a flow to each location $l \in \Loc$ for each variable $v \in \VarInner$.
    \item $\mathit{Sync} = \SyncS ~\cup~ \SyncR$ is a labeling function assigning a set of \emph{sending} and \emph{receiving} variables to each edge  $e \in \Edge$.
    \item \kernel only assigns a distribution for non-receiving edges.
    \item \resetKernel only assigns a distribution for output variables, i.e. $\Psi_{\mathit{var}}^R(\sigma,e)$ is only defined for $\mathit{var} \in \VarInner$ in any state $\sigma$ and any edge $e$.
\end{itemize}
 \end{definition}
The additional labeling function \Sync is used to still adhere to the notation used in \cite{blohm2025towardsQuantitative} but also to clearly distinguish the labels used for the definition of the reset and delay kernel and the labels used to indicate synchronization of edges.
To compose the SHA templates following the \hawksync definition, we have to conservatively extend the rules for \emph{non-synchronized} and  \emph{synchronized} edges from the previously proposed transformation from Simulink to SHA \cite{blohm2025towardsQuantitative} to account for the stochastic kernels.
Thus, the following is only slightly adapted from \cite{blohm2025towardsQuantitative}.
Let us first introduce some notation: Let $\HAWKSet = \{ \hawksyncinstanceI{i} \mid 0 \leq i \leq n\}$ be the set of all SHA templates derived from transforming the Simulink blocks of the model with $\hawksyncinstanceI{i} = (\mathcal{H}_i, \Sync_{i}, \Psi_i, \Psi_i^R)$ and $\mathcal{H}_i=(\Loc_i, \Var_i, \Flow_i, \Inv_i, \Lab_i, \Edge_i,\allowbreak \Init_i)$. $\MappingSet = \{ \MappingI{j} \mid 0 \leq j \leq l \} $ is the set of all synchronization mappings derived from the signal lines of the Simulink model and $\MappingComp = (\variable, \sender, \receiveSet) \in \MappingSet $ is the synchronization mapping that is currently resolved. We refer to the set of SHA templates that is considered in this synchronization mapping as $\HAWKSetSync$. The \intermediateaut considered in this step of the composition  is denoted $\HawkIAInstance$. We refer to the set of SHA templates that are already composed in \HawkIAInstance as $\HAWKSetComposed$ and to the set of all SHA templates that are added in this composition step as $\HAWKSetNew = \HAWKSetSync \backslash \HAWKSetComposed$.


To parallely compose the \intermediateaut \HawkIAInstance with the set of SHA templates \HAWKSetNew via the synchronization mapping \MappingComp = (\variable, \sender, \receiveSet), i.e. $\HawkIAInstanceNew = \HawkIAInstance \parallel_{\MappingComp} \HAWKSetNew$, we now conservatively adapt the definition for \emph{non-synchronized} and  \emph{synchronized} edges from \cite{blohm2025towardsQuantitative} by including the stochastic kernels from the \hawksync as follows:
\begin{definition}
 \label{def:nonsyncjump}
    A non-synchronized edge 
 $j = (\mathbf{l},a,\guard,\reset,\allowbreak\mathbf{l}')$ from source location $\mathbf{l}= (l_1, ..., l_n)$ to target location $\mathbf{l}' = \left(l'_1, ... ,l'_n\right)$ with $l_i, l_i' \in \Loc_i, \Loc_i \in \hawksyncinstanceI{i}$  
exists in the  automaton $\hawksyncinstanceI{\mathit{new}} = \HawkIAInstance \parallel_{\MappingComp} \HAWKSetNew$  iff
\begin{enumerate}
 \item $\exists l_i \in \mathbf{l} \land l'_i \in \mathbf{l}' : \exists j_i \in \Edge_i :  j_{i} =  (l_i,a_i,\guard,\reset,l_i') \land \variable \notin \Sync_i(j_i)$
 , i.e. there exists an edge in one SHA template and that does not send or receive the variable \variable that is synchronized.
    \item $\forall j \neq i : l_j = l_j' $, i.e. no change happens in all other SHA templates.
    \item We assign $\Sync_{\mathit{new}}(j) = \Sync_i(j_i)$.
    \item We assign $\Psi_{\mathit{var}}^R (\sigma, a) = \Psi_{\mathit{var}, i}^R(\sigma_i, a_i) $  for all $\mathit{var} \in\VarInnerI{i} , \sigma \in \Sigma, \sigma_i \in \Sigma_i$, i.e. the reset kernel from the executed edge is assigned to the new edge. 
    \item We assign $\Psi_a(\sigma) = \Psi_{a_i}(\sigma_i)$ for all $\sigma \in \Sigma, \sigma_i \in \Sigma_i$, i.e. the delay kernel from the executed edge is assigned to the new edge. 
\end{enumerate}
 \end{definition}

\begin{definition}
 \label{def:syncjump}
A synchronized edge  $j = (\textbf{l},\guard,\reset,\textbf{l}')$ from a source location $\textbf{l}= (l_1, ..., l_n)$ to a target location $\textbf{l}' = \left(l'_1, ... ,l'_n\right)$ with $l_i, l_i' \in \Loc_i, \Loc_i \in \hawksyncinstanceI{i}$  
exists in the  automaton $\HawkIAInstanceNew = \HawkIAInstance \parallel_{\MappingComp} \HAWKSetNew$  iff 
\begin{enumerate}
    \item $\forall ~l_i \in \mathbf{l},~l_i' \in \mathbf{l}',~\hawksyncinstanceI{i} \in \HAWKSetSync :  \exists j_i \in \Edge_i, ~j_i = (l_i, \guard_i, \reset_i, l_i') ~\land ~\variable \in \Sync(j_i)$, i.e. in all $\hawksyncinstanceI{i} \in \HAWKSetSync$ exists an edge from $l_i$ to $l_i'$ where \variable is either sent or received.    
    \item $\forall l_i \in \mathbf{l},~l_i' \in \mathbf{l'},~\hawksyncinstanceI{i} \notin \HAWKSetSync :  l_j = l_j' $, i.e. no change happens in all $\hawksyncinstanceI{i} \notin \HAWKSetSync$.
    \item $\guard =  \bigwedge_{\substack{1 \leq i \leq n\\ \hawksyncinstanceI{i} \in \HAWKSetSync}} \guard_i, \text{ for }\guard_i \in j_i$, i.e. the guard of $j$ conjoins the guards of $j_i$ in $\hawksyncinstanceI{i} \in \HAWKSetSync$. 
    \item $\reset = \mathit{apply}_{\reset_{\sender}(\variable)} \left(\bigcup_{\substack{1 \leq i \leq n\\ \hawksyncinstanceI{i} \in \HAWKSetSync}} \reset_i \right)  $ where $ \mathit{apply}$ is a function replacing each occurrence of \variable in the reset $r_i$ of the edge $j_i$ in all receiving SHA templates with the linear term $a$, where \hawksyncinstanceI{\sender} is the sending SHA template and $ \reset_{\sender}(\variable) = (\variable {=} a) $. 
    \item $\Sync_{\mathit{new}}(j) = \bigcup_{\substack{1 \leq i \leq n\\ \LACsyncInstanceI{i} \in \LHACSetSync}} \Sync_i(j_i)\backslash \variable $, i.e. we keep all synchronization labels except from the one that is being resolved.
     \item We assign $\Psi_{\mathit{var}}^R (\sigma, a) = \Psi_{\mathit{var}, i}^R(\sigma_i, a_i) $  for all $\hawksyncinstanceI{i} \in \HAWKSetSync, \allowbreak\mathit{var} \in\VarInnerI{i} $, $\sigma \in \Sigma$, $\sigma_i \in \Sigma_i$, i.e. the reset kernels from the executed edges are all assigned to the new edge. 
    \item We assign $\Psi_a(\sigma) = \Psi_{a_{\mathit{send}}}(\sigma_{\mathit{send}})$ for all $\sigma \in \Sigma, \sigma_{\mathit{send}} \in \Sigma_{\mathit{send}}$ where $\hawksyncinstanceI{\mathit{send}}$ is the sending template, i.e. the delay kernel from the sending edge is assigned to the edge. 
\end{enumerate}
 \end{definition}
 After all synchronization mappings are resolved, any undefined reset kernels for a variable $\mathit{var}$ in any state $\sigma$ and edge $a$ will be assigned the identity-remaining Dirac-distribution, i.e. $\Psi_{\mathit{var}}^R(\sigma,a) \sim \mathcal{D}(\mathit{var})$,

\subsection{Additional SHA Templates}
\label{app:templates}
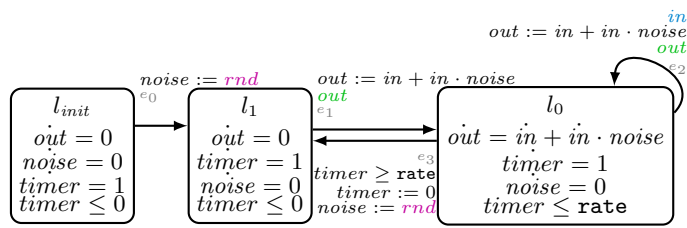
\begin{figure}[b]
    \centering
    \begin{tikzpicture}
    \node[location] (l0) {$l_0$ \\ [\spacer] $\innervar{\dot{\out}}  =\outervar{\dot{\innone}} +\outervar{\dot{\innone}}\cdot \mathit{noise} $ \\ $\innervar{\dot{timer} = 1}$\\ $\dot{\mathit{noise}} = 0$ \\ $timer \leq \simulink{rate}$};
    
    \node[location, left = 1.65cm of l0] (l1) {$l_1$ \\ [\spacer] $\innervar{\dot{\out}}  = 0$ \\ $\innervar{\dot{timer} = 1}$\\$\dot{\mathit{noise}} = 0$\\ $timer \leq 0$};
    
    \node[location, left =  0.7cm of l1] (init) {$l_{\mathit{init}}$ \\ [\spacer] $\innervar{\dot{\out}}  = 0$\\ $\dot{\mathit{noise}} = 0$ \\ $\innervar{\dot{timer} = 1}$ \\ $timer \leq 0$};
     
    \draw[shatransition, sync, bend right = 110, looseness = 2] (l0.20) 
    to node[right,transition,xshift = -2.4cm,yshift=0.32cm, align = right] {
    $\receive{\innone} $ \\
    $\innervar{\out} := \outervar{\innone} + \innone\cdot\innervar{\mathit{noise}} $\\ 
    $\send{\out}$    \\
    {\tiny \color{gray}$\phantom{e^0}e_2$}} (l0.50);

    \draw[shatransition, sync] ([yshift = 0.2cm]l0.west) 
    to node[below,transition, yshift=0.05cm, xshift=0cm,align = right] {        
    {\tiny \color{gray}$\phantom{e^0}e_3$}\\
    $ \innervar{timer} \geq \simulink{rate}$ \\ 
    $\innervar{timer} := 0$ \\
    $\innervar{\mathit{noise}} := \randomclock{rnd}$ 
    }  ([yshift = 0.2cm]l1.east);

    \draw[shatransition, sync] ([yshift = 0.35cm]l1.east)
    to node[above,transition, yshift=0.05cm, xshift = 0.55cm, align = left] {    
    $ \innervar{\out} := \outervar{\innone} + \innone \cdot \innervar{\mathit{noise}} $ \\
    $\send{\out}$    \\[0.5mm]
    {\tiny\color{gray}$e_1$}
    }  
    ([yshift = 0.35cm]l0.west);

       \draw[shatransition, sync] ([yshift= 0.4cm]init.east) to node[above,transition, align = left,yshift=0.2cm, xshift = .5cm] 
       { $\mathit{noise} := \randomclock{rnd}$  \\[0.5mm]
       {\tiny \color{gray}$e_0$}
        } 
     ([yshift= 0.4cm]l1.west);

\end{tikzpicture}
    \caption{SHA Template for Stochastic Noise with Signal as Multiplier.}
    \label{fig:sha-noiseWithIn}
\end{figure}

\begin{figure}[tb]
    \centering
    \begin{tikzpicture}
    \node[location] (aging) {$\mathit{aging}$ \\[\spacer] $\innervar{\dot{\out}}  =\outervar{\dot{\inone}} \cdot (1 - \innervar{age} \cdot \simulink{factor})$\\ $\innervar{\dot{clock}} = 0$\\ $\innervar{\dot{age}} = 0$\\ $\randomclock{\dot{\fail}} = 1 $\\ $\randomclock{\dot{\repair}} = 0 $};    
    \node[location, right=1.2cm of aging] (repairing) {$\mathit{repairing}$ \\[\spacer] $\innervar{\dot{\out}}  =\outervar{\dot{\intwo}} $\\ $\innervar{\dot{clock}} = 0$\\ $\innervar{\dot{age}} = 0$\\ $\randomclock{\dot{\fail}} = 0 $\\ $\randomclock{\dot{\repair}} = 1 $}; 

    \node[location, left = 2.7cm of aging] (fail){$\mathit{fail}$ \\[\spacer] $\innervar{\dot{\out}}  = 0 $\\ $\innervar{\dot{clock}} = 1$\\ $\innervar{\dot{age}} = 0$\\ $\randomclock{\dot{\fail}} = 0 $\\ $\randomclock{\dot{\repair}} = 0 $\\ $\innervar{clock} \leq 0$};

    \draw[shatransition,sync, bend left = 100, looseness = 1.8] (aging.35) 
    to node[transition, xshift= 0.9cm,yshift=0.6cm, align = left] {
    $\receive{\inone} $\\
    $\innervar{\out}:= \outervar{\inone} \cdot (1 - \innervar{age} \cdot \simulink{factor})$\\
    $\send{\out}$\\
     {\tiny\color{gray} $e_0\phantom{e^0}$}
    }  (aging.15);
    
     \draw[shatransition, sync, bend left = 90, looseness = 2] (aging.80) to node[left,transition,xshift = -0.3cm,yshift=-0.2cm] {
    $\receive{\intwo}$ \\{\tiny\color{gray} $e_7$}}  (aging.50);

    \draw[shatransition, sync,bend right = 100, looseness = 1.8] (repairing.40) to node[right,transition,xshift= -1cm,yshift=0.3cm, align  =right] {
    $\receive{\intwo}$\\
    $\innervar{out} := \intwo$\\
    $\send{out}$\\[0.5mm]
    {\tiny\color{gray} $e_1$}
    }  (repairing.70);

    \draw[shatransition, sync,bend left = 100, looseness = 1.8] (repairing.-40) to node[right,transition,xshift= 0cm,yshift=0.3cm, align  =left] {
    {\tiny\color{gray} $e_8$} \\
    $\receive{\inone}$
    }  (repairing.-70);

     \draw[shatransition, sync, bend left = 90, looseness = 2] (fail.85) to node[above,transition,xshift = -0cm,yshift=-0cm, align = left] {
    $\receive{\inone}$ \\[0.5mm]
    {\tiny\color{gray} $e_2$}}  (fail.60);
    
     \draw[shatransition, sync, bend left = 90, looseness = 2] (fail.120) to node[above,transition,xshift = -0cm,yshift=-0cm, align = left] {
    $\receive{\intwo}$ \\[0.5mm]
    {\tiny\color{gray} $e_9$}}  (fail.95);

    \draw[shatransition, sync,] ([yshift=-0cm]repairing.west) to node[below, transition, align = right,xshift=0cm]{
    {\tiny\color{gray} $e_3$}\\
    $\randomclock{\repair}$\\
    $\innervar{age} := 0$\\
    $\innervar{\out} := \outervar{\inone}$\\
    $\send{\out}$
    } ([yshift=-0cm]aging.east);

    \draw[shatransition,sync,] ([yshift=0.37cm]aging.west) to node[below, transition, align= right, xshift = 0.65cm, align = right] {       
    {\tiny\color{gray} $e_4$}\\
    $\randomclock{\fail}$\\
    $\innervar{clock} := 0$
    }  ([yshift=0.37cm]fail.east);

    \draw[shatransition,sync,] ([yshift=0.55cm]fail.east) to node[above, transition, xshift= 0.85cm,yshift=0.cm, align = left] {
    $\innervar{age} < \simulink{steps} - 1$\\
    $\innervar{\out}:= \outervar{\inone} \cdot (1 - (\innervar{age} + 1) \cdot \simulink{factor})$\\
    $\innervar{age} := \innervar{age} + 1$\\
    $\send{\out}$    \\
    {\tiny\color{gray} $e_5\phantom{e^0}$}
    }  ([yshift=0.55cm]aging.west);

    \draw[shatransition,sync] ([yshift=0cm]fail.south) to node[above right, transition,xshift = 0.7cm, yshift = -0.2cm, align= left] {  
    $age \geq \simulink{steps} - 1$\\
    $\innervar{\out} := \outervar{\intwo} $\\
    $\send{\out}$\\
    {\tiny\color{gray} $e_6\phantom{e^0}$}} ([yshift=-0.3cm]fail.south) -| (repairing.south);

\end{tikzpicture}
    \caption{SHA Template for Stochastic Discrete Aging with Explicit Repair Signal.}
    \label{fig:sha-discreteaging2input}
\end{figure}
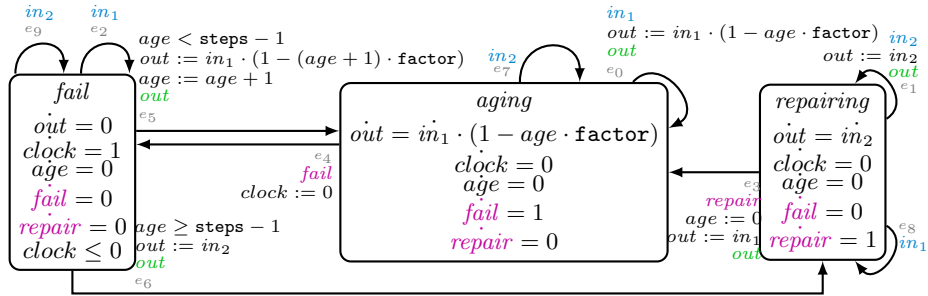

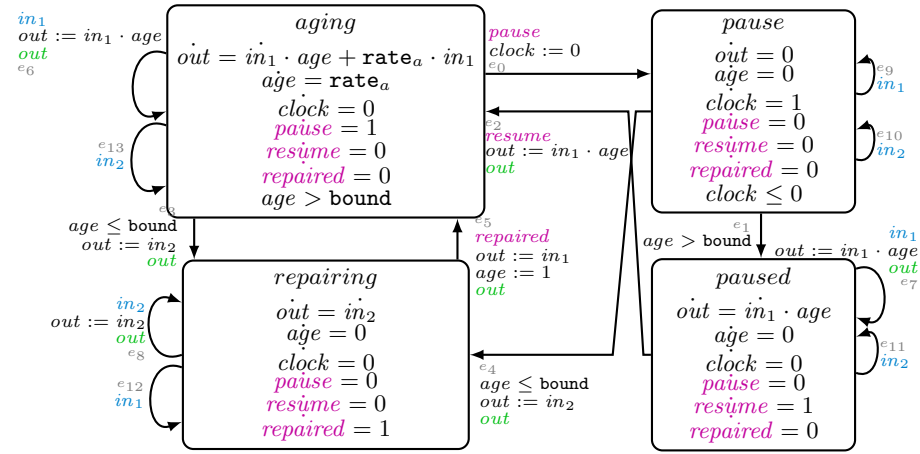
\begin{figure}[tb]
    \centering
    \begin{tikzpicture}
    \node[location, minimum width = 3.8cm] (aging) 
    {$aging$ \\
    [\spacer] $\innervar{\dot{\out}}  =\outervar{\dot{\inone}} \cdot \innervar{\mathit{age}} + \simulink{rate}_a \cdot \outervar{\inone}$ \\ 
    $\innervar{\dot{\mathit{age}}} = \simulink{rate}_a$ \\ 
    $\innervar{\dot{\mathit{clock}}} = 0$\\
    $\randomclock{\dot{\mathit{pause}}} = 1 $ \\ 
    $\randomclock{\dot{\mathit{resume}}} = 0 $ \\ 
    $\randomclock{\dot{\mathit{repaired}}} = 0 $ \\ 
    $\innervar{\mathit{age}} > \simulink{bound}$ 
    };
    
    \node[location, below=4ex  of aging, minimum width = 3.8cm] (repairing) 
    {$repairing$ \\
    [\spacer] $\innervar{\dot{\out}}  = \dot{\intwo}$ \\
    $\innervar{\dot{\mathit{age}}} = 0$ \\ 
    $\innervar{\dot{\mathit{clock}}} = 0$\\
    $\randomclock{\dot{\mathit{pause}}} = 0 $ \\
    $\randomclock{\dot{\mathit{resume}}} = 0 $ \\
    $\randomclock{\dot{\mathit{repaired}}} = 1 $
    };
    
    \node[location,right=2.2cm of aging, minimum width = 2.7cm] (pause)
    {$pause$ \\ 
    [\spacer] $\innervar{\dot{\out}}  = 0$ \\ $\innervar{\dot{\mathit{age}}} = 0$ \\ 
    $\innervar{\dot{\mathit{clock}}} = 1$\\
    $\randomclock{\dot{\mathit{pause}}} = 0 $ \\ 
    $\randomclock{\dot{\mathit{resume}}} = 0 $ \\ 
    $\randomclock{\dot{\mathit{repaired}}} = 0 $\\
    $\innervar{\mathit{clock}} \leq 0$
    };
    
    \node[location, minimum width = 2.7cm] (paused) at (pause|-repairing)
    {$paused$ \\ 
    [\spacer] $\innervar{\dot{\out}}  = \innervar{\dot{\inone} \cdot \innervar{\mathit{age}}}$ \\ $\innervar{\dot{\mathit{age}}} = 0$ \\ 
    $\innervar{\dot{\mathit{clock}}} = 0$\\
    $\randomclock{\dot{\mathit{pause}}} = 0 $ \\ 
    $\randomclock{\dot{\mathit{resume}}} = 1 $ \\ 
    $\randomclock{\dot{\mathit{repaired}}} = 0 $
    };

     \draw[shatransition,sync,] ([yshift=0.5cm]aging.east) to node[above,yshift = -0.1cm,xshift =-0.45cm, transition, align = left] 
     {
     $\randomclock{\mathit{pause}}$ \\
     $\innervar{\mathit{clock}} := 0$ \\[0.5mm]
      {\tiny \color{gray} $e_0$}
     }  ([yshift=0.5cm]pause.west);

    \draw[shatransition,sync,] ([xshift=0.1cm]pause.south) to node[left, transition, align = right] 
    { 
     {\tiny \color{gray} $e_1$}\\$\innervar{\mathit{age}} > \simulink{bound}$
    }  ([xshift=0.1cm]paused.north);
    
    \draw[shatransition, sync,] ([yshift=-0cm]paused.west) -- ([xshift=-0.1cm]paused.west) -- 
    ([xshift = 1.9cm,yshift=0cm]aging.east) --  node[below, transition, align = left, xshift = 0cm] {
     {\tiny \color{gray} $e_2$}\\$\randomclock{\mathit{resume}}$ \\
    $\innervar{\out} := \outervar{\inone} \cdot \innervar{\mathit{age}}$ \\
    $\send{\out}$
    } ([yshift=0cm]aging.east);
    
     \draw[shatransition, sync,] ([xshift=-1.75cm]aging.south) to node[left, transition, align = right, near start, yshift=-1mm, xshift = -0.1cm] 
     {
     {\tiny \color{gray} $e_3$}\\
     $\innervar{\mathit{age}} \leq \simulink{bound}$\\
     $\innervar{\out} := \intwo$\\
    $\send{\out}$
     }  ([xshift=-01.75cm]repairing.north);
     
     \draw[shatransition, sync,] ([xshift=0cm]pause.west) -- 
      ([xshift=-0.2cm]pause.west)
     --
      ([xshift=1.9cm]repairing.east) --node[ below,transition, align = left,xshift = -0.1cm, yshift = 0cm] 
     {
     {\tiny \color{gray} $e_4$}\\
     $\innervar{\mathit{age}} \leq \simulink{bound}$\\
     $\innervar{\out} := \intwo$ \\
    $\send{\out}$
     }  (repairing.east);
     
    \draw[shatransition, sync,] ([xshift=1.75cm]repairing.north) to node[xshift =0.1cm,yshift = -0.2cm,right, transition, align= left] {
    {\tiny \color{gray} $e_5\phantom{e^0}$}\\
     $\randomclock{\mathit{repaired}}$ \\ 
    $\innervar{\out} := \outervar{\inone}$ \\
    $\innervar{\mathit{age}} := 1$\\
    $\send{\out}$\\
    }  ([xshift=1.75cm]aging.south);

    \draw[shatransition, sync, bend right = 110, looseness = 2] (aging.160) to node[near start,transition,xshift = -0.7cm,yshift=0.2cm, align= left] {
     $\receive{\inone} $ \\
     $\innervar{\out} := \outervar{\inone} \cdot \innervar{\mathit{age}} $   \\
     $\send{\out}$\\[0.5mm]
     {\tiny \color{gray} $e_6$}}  (aging.180);
    
    \draw[shatransition,sync, bend left = 110, looseness = 2] (paused.40) to node[near start, transition, xshift=-0.4cm,yshift=0.2cm, align = right]
    {
    $\receive{\inone}$\\
    $\innervar{\out} := \outervar{\inone} \cdot \innervar{\mathit{age}} $\\ 
    $\send{\out}$\\[0.5mm]
    {\tiny \color{gray} $e_7$}
    }  (paused.20);
    
    \draw[shatransition,sync, bend left = 110, looseness = 2.2] (repairing.180) to node[near start,transition,xshift = -0.8cm,yshift=0.3cm, align= right]
    {
    $\receive{\intwo}$\\
    $\innervar{\out} := \intwo $\\ 
    $\send{\out}$\\[0.5mm]
    {\tiny \color{gray} $e_8$}
    }  (repairing.160);

    \draw[shatransition,sync, bend right = 110, looseness = 2] (pause.10) to node[transition, xshift= 0.25cm,yshift=-0cm, align = left] {
     {\tiny \color{gray} $e_9$}\\$\receive{\inone} $}  (pause.25);

    \draw[shatransition,sync, bend right = 110, looseness = 2] (pause.-25) to node[transition, xshift= 0.25cm,yshift=-0cm, align = left] {
     {\tiny \color{gray} $e_{10}$}\\$\receive{\intwo} $}  (pause.-10);

    \draw[shatransition,sync, bend right = 110, looseness = 2] (paused.-10) to node[transition, xshift= 0.25cm,yshift=-0cm, align = left] {
     {\tiny \color{gray} $e_{11}$}\\$\receive{\intwo} $}  (paused.10);

    \draw[shatransition,sync, bend right = 110, looseness = 2] (repairing.185) to node[transition, xshift= -0.3cm,yshift=-0cm, align = right] {
     {\tiny \color{gray} $e_{12}$}\\$\receive{\inone} $}  (repairing.205);
     
    \draw[shatransition,sync, bend right = 110, looseness = 2] (aging.185) to node[transition, xshift= -0.3cm,yshift=-0cm, align = right] {
     {\tiny \color{gray} $e_{13}$}\\$\receive{\intwo} $}  (aging.205);
\end{tikzpicture}
    \caption{SHA Template for Stochastic Continuous Aging with Explicit Repair Signal.}
    \label{fig:sha-continuousteaging2input}
\end{figure}

\paragraph{SHA Template for Noise} The SHA template shown in \Cref{fig:sha-noiseWithIn} formalizes the Simulink subsystem given in \Cref{fig:simulink-noise} when using the input signal as a multiplier for the distortion. The main difference to the SHA template with a constant factor (see \Cref{fig:sha-noise} is that the value of \out is now assigned to $\out := \innone + \innone\cdot \mathit{noise}$ and the flow of \out is given by $\dot{\out} = \dot{\innone} + \dot{\innone}\cdot\mathit{noise}$. 
The reset kernel is given by $\Psi^R_{\mathit{noise}} (\sigma,e_j)\sim \mathcal{N}(\simulink{mean},\simulink{var})  $ for $j \in \{0,3\} $ and all states $\sigma$.
The initial state is given by $\Init(l_{\mathit{init}}) = (\{\out= 0,\mathit{timer} := 0,\mathit{noise} := 0\},  \texttt{true})$.
\paragraph{SHA Template for Discrete Aging} The SHA template shown in \Cref{fig:sha-discreteaging2input} formalizes the Simulink subsystem given in \Cref{fig:simulink-agingdiscrete} for the case when an explicit input signal for the output when repairing is used. The main difference to the SHA template where a constant repair rate $\simulink{rate}_r$ of the regular input signal is used (see \Cref{fig:sha-discreteaging}) is that the flow of \out in location $\mathit{repairing}$ now equals the flow of \intwo and the value of \out is assigned to the value of \intwo at the edge to $\mathit{repairing}$. Additionally, discrete updates of \intwo are handled at self-loops.
The delay kernel is defined as $\Psi_{e_3} (\sigma)\sim \simulink{Dist}_1$,$\Psi_{e_4}(\sigma)\sim \simulink{Dist}_2$ with $ \simulink{Dist}_i=\mathcal{U}(\simulink{low}, \simulink{high}) \mid \mathcal{N}_{\geq 0} ~(\simulink{var}) $  for $i \in\{1,2\}$ and all states $\sigma$.
The initial state is given by $\Init(l_{\mathit{aging}}) = (\{\out= \innone,\randomclock{\mathit{fail}} := 0,\randomclock{\mathit{repair}} := 0, \mathit{age} = 0, \mathit{clock} = 0\}, \texttt{true})$.
\paragraph{SHA Template for Continuous Aging} The SHA template shown in \Cref{fig:sha-continuousteaging2input} formalizes the Simulink subsystem given in \Cref{fig:simulink-agingcontinuous} for the case when an explicit input signal for the output when repairing is used.  The main difference to the SHA template where a constant repair rate $\simulink{rate}_r$ of the regular input signal is used (see \Cref{fig:sha-continuousaging}) is the adjusted flow and resets of the variable \out, analogously to the changes for the discrete aging template with the explicit repair signal.
The delay kernel is defined as $\Psi_{e_0} (\sigma)\sim \simulink{Dist}_1$, $\Psi_{e_2}(\sigma)\sim \simulink{Dist}_2$, $\Psi_{e_5}(\sigma)\sim \simulink{Dist}_3$ with $ \simulink{Dist}_i=\mathcal{U}(\simulink{low}, \simulink{high}) \mid \mathcal{N}_{\geq 0} ~(\simulink{var}) $ for $i \in\{1,2,3\}$ and all states $\sigma$.
The initial state is given by $\Init(l_{\mathit{aging}}) = (\{\out= \innone\cdot \mathit{age},\randomclock{\mathit{pause}} := 0,\randomclock{\mathit{resume}} := 0,\randomclock{\mathit{repair}} := 0, \mathit{age} = 1, \mathit{clock} = 0\}, \texttt{true})$.

\FloatBarrier

%
%
\bibliographystyle{splncs04}
\bibliography{lit}
%

\end{document}